\documentclass[manuscript]{acmart}
\usepackage{natbib}
\usepackage{colortbl}
\usepackage{multirow}
\usepackage{graphicx}
\usepackage{subfig}

\usepackage{comment}
\usepackage{longtable}
\usepackage{tipa}
\usepackage{soul}
\usepackage{pgfplots}
\usepackage{multirow}

\AtBeginDocument{%
  \providecommand\BibTeX{{%
    \normalfont B\kern-0.5em{\scshape i\kern-0.25em b}\kern-0.8em\TeX}}}
\acmYear{2024}\copyrightyear{2024}
\setcopyright{rightsretained}
\acmDOI{10.1145/3613904.3642605}
\acmISBN{979-8-4007-0330-0/24/05}

\begin{document}

\title{Low-resourced Languages and Online Knowledge Repositories: A Need-Finding Study}


\author{Hellina Hailu Nigatu}
\affiliation{ \institution{UC Berkeley} 
\country{USA}}
  \email{hellina_nigatu@berkeley.edu}

\author{John Canny}
\affiliation{%
  \institution{UC Berkeley}
  \country{USA}}
  \email{canny@berkeley.edu}

\author{Sarah E. Chasins}
\affiliation{%
  \institution{UC Berkeley}
   \country{USA}}
  \email{schasins@eecs.berkeley.edu}


\begin{abstract}

 Online Knowledge Repositories (OKRs) like Wikipedia offer communities a way to share and preserve information about themselves and their ways of living. However, for communities with low-resourced languages---including most African communities---the quality and volume of content available are often inadequate. One reason for this lack of adequate content could be that many OKRs embody Western ways of knowledge preservation and sharing, requiring many low-resourced language communities to adapt to new interactions. To understand the challenges faced by low-resourced language contributors on the popular OKR Wikipedia, we conducted (1)~ a thematic analysis of  Wikipedia forum discussions and (2)~ a contextual inquiry study with 14 novice contributors. We focused on three Ethiopian languages: Afan Oromo, Amharic, and Tigrinya. Our analysis revealed several recurring themes; for example, contributors struggle to find resources to corroborate their articles in low-resourced languages, and language technology support, like translation systems and spellcheck, result in several errors that waste contributors' time. We hope our study will support designers in making online knowledge repositories accessible to low-resourced language speakers. 

\end{abstract}



\begin{teaserfigure}
    \subfloat[Amharic article about the Big Mac burger.]{\includegraphics[width=0.5\textwidth]{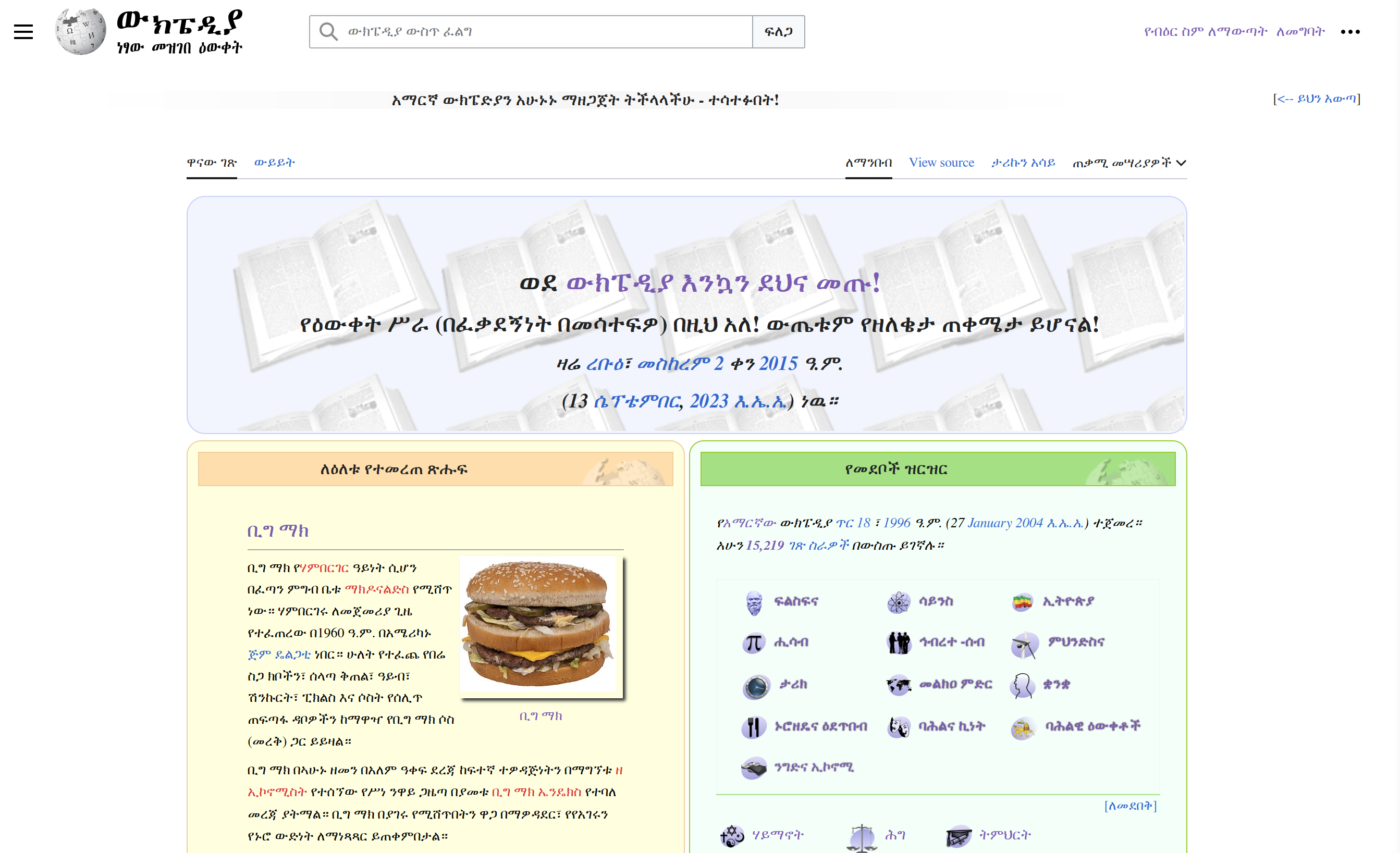}\label{fig:big_mac}}
   \subfloat[English article about a famous community in Ethiopia.]{\includegraphics[width=0.5\textwidth]{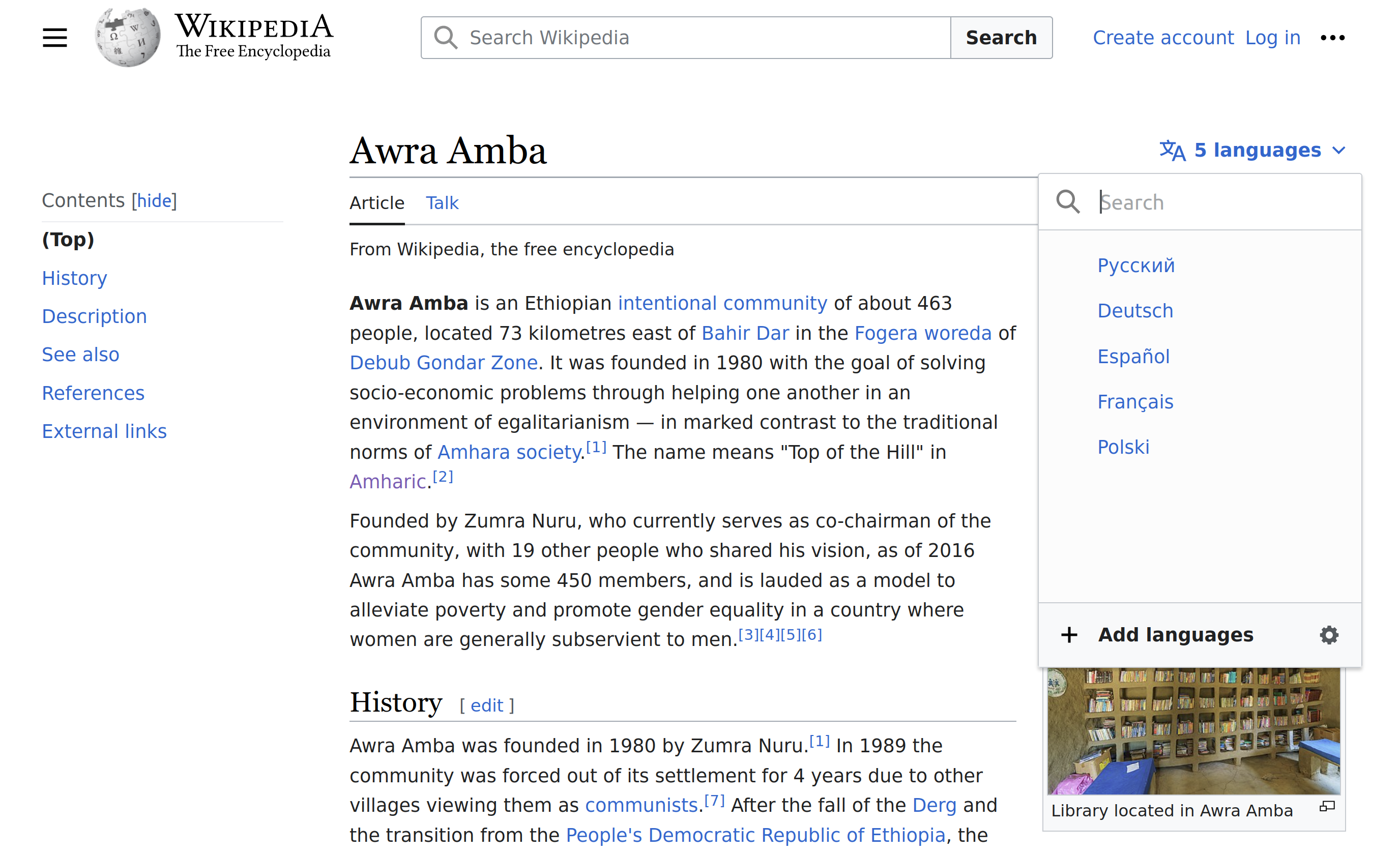}\label{fig:awra_amba}}
  \caption{Fig. \ref{fig:big_mac} shows an article on the main page of the Amharic Wikipedia about the Big Mac burger even though there are no McDonald's restaurants in Ethiopia. On the other hand, an article about a famous Ethiopian community, Awra Amba (Fig. \ref{fig:awra_amba}), is available in five languages on Wikipedia, none of which are Ethiopian languages.}
  \label{fig:teaser}
\end{teaserfigure}

\maketitle
\section{Introduction} \label{intro}

Knowledge repositories are a means to preserve and share information about communities, cultures, and other expertise across the world. \emph{Online} knowledge repositories allow for the dissemination of knowledge across borders with a wide set of users. One such platform is Wikipedia, an open-source online knowledge repository available in 332 languages and with over 191 million articles~\cite{wiki_list_2023}. According to its foundation's vision statement, Wikipedia is a community-rooted effort with the goal to create a world where ``\ldots{}every single person on the planet is given free access to the sum of all human knowledge''~\cite{noauthor_wikipedia_2004}. But while Wikipedia is available in over 300 languages, not all languages are evenly represented. In particular, the representation of African communities who mostly speak low-resourced languages---languages that do not have a lot of data available online~\cite{nekoto_participatory_2020}---lags in quantity and quality.

In this paper, we focus on languages spoken in Ethiopia, a country with over 120 million people~\cite{pop_ethiopia_2023}. In particular, we focus on three of the four most-spoken first languages in Ethiopia: Afan Oromo, Amharic, and Tigrinya~\cite{twb_language_2021}. There are 6.6 million articles in English-language Wikipedia compared to 15,190 articles in Amharic, 1,258 articles in Afan Oromo, and 257 articles in Tigrinya~\cite{wiki_list_2022}. In addition to the small quantity, many articles in these languages are \emph{stubs} (articles that are marked as too short or incomplete), written in a different language other than the language of the Wikipedia, or have other quality issues such as poor spelling. Finally, articles in these languages lack contextual relevance; many articles written in the languages of the communities are about topics unrelated to the community, while articles about the community may not be accessible in their own language or languages (Fig.~\ref{fig:teaser}). The lack of good quality content in low-resourced languages also trickles down to language technologies built using this content. Wikipedia has been used as a source to train NLP models~\cite{dodge_documenting_2021, kalyan_ammus_2021}, but our work reveals that in the context of low-resourced languages, it exhibits quality, quantity, and relevance issues, and even includes abusive content. Previous literature in NLP has also criticized large crawled datasets (like Wikipedia) for lacking contextual relevance in the context of African communities~\cite{ogueji_small_2021}. Using Wikipedia for training means the issues with low-resourced language Wikipedia can ripple out into other domains and cause downstream harm.

Decolonial scholars have called for African literature to be written in African languages~\cite{bhola_ngugi_1987, tamale_decolonization_2020}. Extending this argument, we posit it is valuable for African knowledge repository content to be written in African languages. However, technological interventions to achieve this goal may be risky. Previous work has shown that developing technologies for preserving Indigenous Knowledge (IK) is challenging---in particular, it can compromise the agency of knowledge creators~\cite{kotut_clash_2020}. Nevertheless, although we stress that we must not expose communities to exploitation via technological intervention, it is clear that currently available OKRs do not support communities to preserve their knowledge in their languages and in their own ways~\cite{gallert_indigenous_2016}.  This implicitly bars some communities from interacting with OKRs. We see mounting evidence of excitement to grow the African Wikipedia. For instance, African Wikipedians (Wikipedia contributors) gather at WikiIndaba \footnote{``indaba'' is a Zulu word which means ``gathering'' or ``meeting.''}, a conference in which participants deliberate over ways to increase ``coverage and involvement of Africa in Wikimedia projects''~\cite{indaba_wikiindaba_2023}. In this work, we ask: \textit{What challenges do low-resourced language contributors face when interacting with current OKRs?}  


To understand the obstacles to contributing to OKRs in low-resourced languages, we framed two key research goals: First, we wanted to understand what challenges existing, experienced Wikipedia contributors faced. However, the number of contributors for these languages is low (see Section \ref{background}). Hence, our second goal was to investigate what barriers stand between novices when they try to make Wikipedia contributions in their languages. Thus, we conducted two studies: (1)~To meet our first goal, we collected and analyzed forum discussions among experienced Wikipedia contributors in our three study languages.  (2)~To meet our second goal, we conducted a contextual inquiry study with \textit{novice} Wikipedia contributors in the same three languages. Our studies revealed how (i)~Wikipedia's design and (ii)~the failures of key language technologies (e.g., machine translation, search) obstruct contributors.  
We also found that, beyond technological barriers, contributors are affected by the availability of scholarly work and financial resources.

We make two primary contributions: First, we contribute an analysis of both Wiki forum data and our contextual inquiry sessions; these analyses broaden the HCI4D community's understanding of the challenges faced by low-resourced language contributors interacting with OKRs and contribute to the growing body of HCI literature~\cite{kotut_clash_2020, alvarado_garcia_decolonial_2021, pendse_treatment_2022, das_decolonization_2023, ali_brief_2016} on the decolonial design of technologies for marginalized communities. Second, based on these analyses, we propose a set of design opportunities for researchers and designers who want to build inclusive language technologies.
By making technologies accessible to low-resourced language speakers, we can empower communities to share information about their own cultures in their own languages; we believe it should be communities' own choices---not design barriers---that determine whether or not communities share information in OKRs.

We structure our paper as follows: First, in Section \ref{related_work}, we provide a literature review to situate our work within the existing HCI and decolonial studies literature, as well as literature on improving Wikipedia for different languages. Section \ref{background} describes the linguistic and cultural aspects of the three languages under study, which will be necessary for understanding our findings. In the same section, we provide key background information about the quality and volume of content in the three languages' Wikipedias. In Section \ref{forum_analysis}, we describe the methodology and findings of our forum analysis, followed by Section \ref{contextual_inquery} in which we present the methodology and findings of our contextual inquiry. Then, in Section \ref{limitation}, we discuss the limitations of our work. Section \ref{lessons}
presents design opportunities and offers directions for future research. In Section  \ref{disscussion}, we discuss challenges in designing for such communities and reflect upon our own practices. Section \ref{conclusion} concludes.

\section{Related Work} \label{related_work}
 In this section, we discuss prior work on the interaction between IK and Technology, work on designing for marginalized communities, and previous interventions aimed at closing the gap in Wikipedia articles across languages.  
      
\subsection{Indigenous Knowledge Preservation and Technology} 
       Studies~\cite{giglitto_eye_2018, lu_i_2019, liu_generating_2022} focusing on Intangible Cultural Heritage (ICH) have shown how several indigenous communities preserve IK through practices like oral history, performance art, rituals and festivals. The HCI community has also studied how technology plays a role in the preservation of IK: for instance, video communication can help people far from their ancestral lands stay connected to IK systems ~\cite{awori_transnationalism_2015, taylor_preserving_2018, awori_sessions_2016}. Social media platforms and virtual reality can also help local communities consume and preserve IK~\cite{siew_participatory_2013, kotut_trail_2021, kotut_winds_2022, skovfoged_tales_2018, smith_preserving_2018}. However, technological advances do not always match IK preservation methods~\cite{bidwell_pushing_2011, kotut_trail_2021}, which usually involves multi-modal interactions---like oral traditions and rituals---and differs from one community to another. Gallert et al.~\cite{gallert_indigenous_2016} criticized the text-only interaction mode of online knowledge repositories for excluding oral tradition-based knowledge preservation systems. Beyond input modalities, Gallert et al.~\cite{gallert_indigenous_2016} also criticized the citation and editing features of Wikipedia, showing how Western-centred editing and participation rules inhibit participation on the platform.
       
    Issues with integrating technology into IK preservation mechanisms are not limited to the divergence in ways of knowledge preservation. Previous work~\cite{graham_uneven_2014,graham_augmented_2013, van_pinxteren_african_2017} provides evidence that, despite the hopes of democratization through internet connectivity, there is a geographical divide in information representation. One study ~\cite{van_pinxteren_african_2017} found that only 1\% of all Wikipedia articles were in African languages. Other studies~\cite{wyche_deliberate_2010, wyche_facebook_2013} have found that high cost and access to infrastructure also contribute to patterns of online platform usage in Global South communities. In addition to connectivity, Graham et al.~\cite{graham_uneven_2014} argue that issues such as access to reference materials could constrain users from contributing content in local languages. 

       While verifiable sources and notability are important, it may also make it difficult to create local, contextual content about certain communities. Previous studies~\cite{maclure_overlooked_2006, mitchell_education_2020, waruru_renowned_2022, abebe_narratives_2021} show how the global research community has ignored huge bodies of work by African researchers. Other work~\cite{kotut_winds_2022, tamale_decolonization_2020, kessi_decolonizing_2020} have also shown that community members are skeptical of content written by outsiders, usually colonialists. This trend of authorship about African communities coming predominantly from outside of the communities is not limited to research papers; most Wikipedia article contributions about African communities are not made from within the continent~\cite{gallert_indigenous_2016}. Our work focuses exclusively on understanding IK contributors and tries to answer the question: What challenges prevent such contributors from creating content in their languages on OKRs?

\subsection{Designing Technologies for Marginalized Communities} \label{related_work_harms}

          Given the low-resourced language audiences we intend to support, our work necessarily builds on the prior literature on designing technologies for marginalized communities.
          While technological barriers alone may explain the data gap on OKRs, it is also possible that communities \emph{choose} to withhold their content from online repositories. Abebe et al.~\cite{abebe_narratives_2021} shed light on how trust, colonial legacies, and exploitation, among other issues, shape the data-sharing landscape in Africa. Low-resourced language speaking communities that come from marginalized groups in the Global South have historically been excluded from much of the research in the design and study of technology~\cite{linxen_how_2021, septiandri_weird_2023}. Technology may be a double-edged sword for communities that have historically been exploited through colonization, as technology could put them at risk of automated exploitation~\cite{birhane_algorithmic_2020, alkhatib_live_2021, benjamin_race_2020}. 
          Concerning IK, communities may not want all and every aspect of their practices to be available online. Previous work~\cite{kotut_winds_2022} discusses the challenges of balancing design for keeping sensitive IK private (only for the intended audience) while also supporting community members who seek such knowledge but lack direct access.   
          
          However, the risk of technological exploitation does not necessarily mean we should leave communities behind altogether under the guise of protecting them. As Benjamin \cite{benjamin_race_2020} argues, one of the ways technology further marginalizes groups is by excluding them in ``design, intent, implementation, and results.'' The HCI community has approached the sensitive topic of designing for marginalized groups through critical race theory ~\cite{ogbonnaya-ogburu_critical_2020, chen_collecting_2022, smith_whats_2020}, decolonial \cite{alvarado_garcia_decolonial_2021, lazem_challenges_2022, ali_brief_2016} and post-colonial\cite{irani_postcolonial_2010, ahmed_residual_2015} lenses and feminist ~\cite{okerlund_feminist_2021, sondergaard_feminist_2022, spors_selling_2021} lenses. Previous works~\cite{le_dantec_strangers_2015, harrington_engaging_2019} have offered examples of how co-design and participatory design approaches can ensure community values are centered in design. Specific to IK, scholars~\cite{kotut_clash_2020, sheehan_indigenous_2011} have presented frameworks that center on respecting community members' agency and protecting them from harm.




\subsection{Improving Wikipedia for Low-Resourced Languages}

Previous works~\cite{team_no_nodate, hadgu_lesan_2021} have proposed to increase Wikipedia content in low-resourced languages via the use of Machine Translation (MT) systems. While this approach can increase the number of articles in each Wikipedia, researchers and the Wikipedia community have raised concerns that: (1) simply putting the article through MT systems is redundant, as one can use the translation feature on the Wikipedia interface to read the article in a given language \cite{meta_machine_2023}; (2) MT systems still have significant performance gaps for low-resourced languages \cite{nekoto_participatory_2020, emezue_mmtafrica_nodate, hadgu_lesan_2021}; and (3) (machine) translated articles have a risk of lacking contextual relevance to communities\cite{wulczyn_growing_2016}. Wikipedia's own Machine Translation toolkit guideline \cite{meta_machine_2023} states ``an unedited machine translation, left as a Wikipedia article, is worse than nothing.'' 

Alternative approaches include tools for suggesting topics.  For example, prior work developed a system that helps contributors identify new articles to create based on articles that are found in other languages~\cite{wulczyn_growing_2016}; this work avoids suggesting articles that do not have cultural significance by using a regression model to rank its suggestions, based on features like page view counts in other languages, number of Wikipedias that include the article, and geographic location of viewers. Another work used visualizations to show potential contributors what concepts are not represented in each language~\cite{miquel-ribe_wikipedia_2020}. Our work differs in that (1) we take a step back from assumptions about potential solutions and study the contributors themselves and (2) we focus not on topic selection but on the writing process, be it with the use of MT or not. 

\section{Background} \label{background}

In this section, we provide information to contextualize and understand our findings through (1)~background on the Ethiopian languages at the center of our study and (2)~data on the existing Wikipedia articles in these languages.

\subsection{Languages of Study and the Ethiopian Calendar}
Here we summarize important properties of the languages in our study, and we describe the Ethiopian calendar to give more context to our findings.  
 
 \paragraph{Languages} Two of the languages in our study, Tigrinya and Amharic are Afro-Semitic languages written in the Ge'ez script. The Ge'ez script follows an Abugidas/Syllabic alphabet \cite{abugidas_nodate}, meaning it has symbols for consonants and vowels and each consonant has a vowel which can be changed to another vowel or muted by diacritics or modifications. When you bring a vowel and a consonant together, it can either change the symbol of the consonant or be written as two separate characters. Some characters present in one language may not be present in the other. Additionally, identical letters might have different sounds across languages: for example, in the Amharic language there are 4 different letters for the glottal stop sound ``\textglotstop a'' while in Tigrinya those four letters each have a distinct sound: ``\textglotstop a'', ``\textglotstop \textturna'', ``\textrevglotstop a'', and ``\textrevglotstop \textturna'' \cite{ipa_amharic_nodate, ipa_tigrinya_2021}. Hence, there might be spelling differences in Amharic such that the pronunciation and meaning are the same but the word can be spelled using any of the multiple different letters that represent the sound. 
 The third language, Afan Oromo, is a Cushtic language with the Qubee alphabet, written with Latin characters~\cite{omniglot_oromo_nodate}. In Afan Oromo, while different dialects may have different spellings, spelling differences as small as a single letter could also result in entire meaning changes: for instance, ``coora'' means ``feeling'', ``cora'' means ``gathering'', and ``coraa'' means ``remnant''~\cite{omniglot_oromo_nodate}.

\paragraph{Calendar} The Ethiopian calendar system has 13 months in total. The first 12 months each have 30 days, and the last month has 5 or 6 days, depending on leap years. In writing, dates are usually marked with abbreviations to indicate the calendar system: Gregorian Calendar dates are marked with ``ende awropawian aqotater (e.ae.a)'' \footnote{We could not add Ge'ez characters with the currently supported TAPS packages; hence, we resorted to providing the Romanized writing for the Amharic words.} in Amharic or ``Akka Lakkoofsa Awurooppaa (A.L.A)'' in Afan Oromo. Both translate to ``According to the European Calendar.''

\subsection{Quantity and Quality of Wikipedia Articles} \label{issues}

Here we contextualize the issues with existing articles in Afan Oromo, Tigrinya, and Amharic. This background (1)~provides context for the types of issues we observed in our two studies (Section \ref{forum_analysis} and Section \ref{contextual_inquery}) and (2)~offers further evidence of \textit{why} this problem domain is important. 
The information in this background section represents statistics about Wikipedia for all three study languages.
As a point of comparison, we also include data about the Arabic and English Wikipedias. Arabic is higher-resourced than our study languages but lower-resourced than English; Arabic is also written in its own non-Latin script.

\subsubsection{Quantity: There are Limited Number of Articles in Low-Resourced Languages.}

\begin{table*}[]
    \centering
    \begin{tabular}{|c|p{1.5cm}|p{2cm}|p{1.5cm}|c|p{1.5cm}|p{2cm}|}
    \hline
         \textbf{Language} & \textbf{Article Count} & \textbf{New Articles per Day} & \textbf{Edits Per Month} & \textbf{Active Editors} & \textbf{Speakers (in Millions)} & \textbf{Editors per Million speakers} \\
    \hline
         Tigrinya & 304 & 0 & 1 & 0 & 6.7 & 0.0\\
         Oromo & 1,048 & 0 & 63 & 2 & 25.5 & 0.1\\
         Amharic & 15,039 & 1 & 798 & 7 & 25 & 0.3\\
    \hline
        Arabic & 653,195 & 614 & 409,307 & 823 & 422 & 2 \\
    
         English & 5,779,516 & 634 & 3,256,712 	& 30,205& 1,121 	& 27 \\

    \hline
    \end{tabular}
    \caption{\textbf{Official Statistics from Wikipedia about five languages' Wikipedia.\cite{stats_wikimedia_2023}} We have added the stats for English and Arabic Wikipedia as a way to calibrate the extent to which the three low-resourced languages' Wikipedia is lagging in terms of quantity. For instance, there are 27 editors per million speakers for English compared to less than 1 editor per million speakers for any of the three languages.}
    \label{tab:stats}
\end{table*}

  Table \ref{tab:stats} shows the number of articles for the three languages is in the tens of thousands, thousands, and hundreds. Additionally, we see fewer than 1 editor per million speakers for Afan Oromo, Tigrinya, and Amharic. We observe that the number of active editors is less than 10 for all three languages, with Tigrinya at zero active editors.  See Table \ref{tab:stats} for a comparison with Arabic and English.

 \begin{figure}
     \centering
     \includegraphics[width=0.5\textwidth]{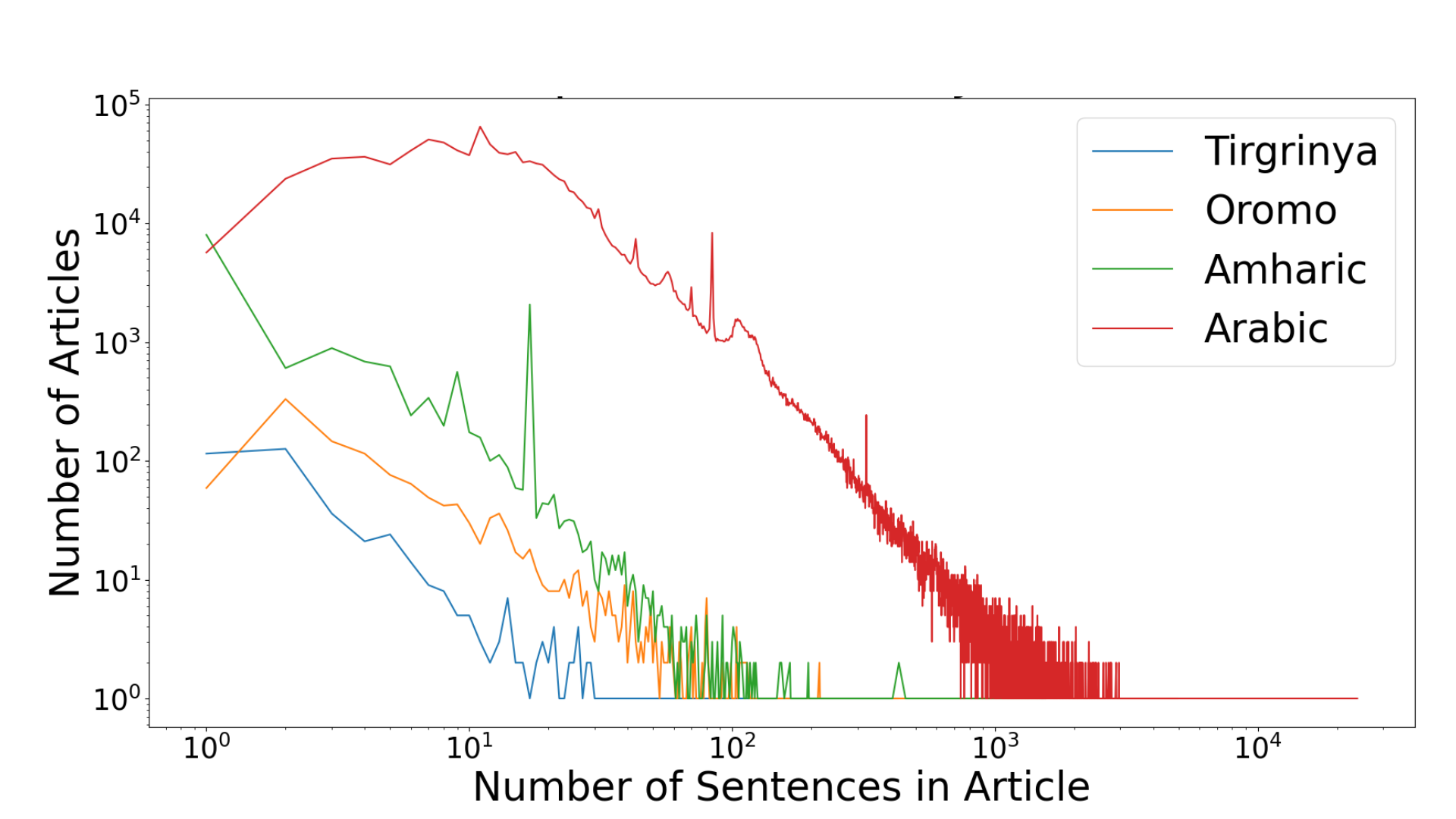}
     \caption{\textbf{Plot showing the distribution of sentence count per article for Amharic, Tigrinya, Afan Oromo, and Arabic Wikipedia.} We observe that for Amharic, Tigrinya, and Afan Oromo, the distributions have a spike at the beginning, with most articles having only one or two sentences. On the other hand, the distribution for the Arabic data shows a spike at 10 sentences. We also observe a concentration in the middle of the distributions, with some articles having from 20 to 200 sentences for Afan Oromo and Amharic and from 10 to 30 sentences for Tigrinya. For Arabic, the distribution is shifted to the right, with articles in the thousands of sentence count ranges having a dense distribution. We observe a plateau for all Wikipedia with a higher number of sentences only having a single article count.}
     \label{fig:sentence_count}
 \end{figure}

\subsubsection{Quality: Articles may be Stubs, Link to Non-Existing Pages, or be in Another Language.}

To analyze article quality, we used dumps from Wikipedia Downloads \cite{dumps_wikimedia_2023} for all three study languages, which includes all articles plus discussion for each Wikipedia.
  The Tigrinya Wikipedia data included 709 titles, Afan Oromo 2,180 titles, and Amharic 22,887 titles.  
  For each of the three study languages, the Wikipedia data dumps included titles without associated article text---14/709 in Tigrinya, 161/2,180 in Afan Oromo, and 79/22,887 in Amharic.
  Each language also included duplicate article texts---277/709 in Tigrinya, 415/2,180 in Afan Oromo, and 6,745/22,887 in Amharic which we excluded when producing the statistics in the next paragraphs.

Although length is not a perfect proxy for quality, we first ask what proportion of articles are very short.  Fig. \ref{fig:sentence_count} shows sentence count for articles in Amharic, Afan Oromo, Tigrinya, and, for comparison, Arabic. We show the log-log scale distribution of sentence counts per article. A deeper look at the list of article titles for Amharic Wikipedia revealed articles include  2 and sometimes 3 stubs for each of the years in the Gregorian calendar (Section \ref{background}). Stubs were created with two different spellings for `Gregorian Calendar' or omitting `Gregorian Calendar' and only writing the year resulting in two or three stubs per year; together these stubs account for thousands of the 15,190 articles in the Amharic data dump.

\begin{figure}[!tbp]
          \centering
          \includegraphics[width=0.45\textwidth]{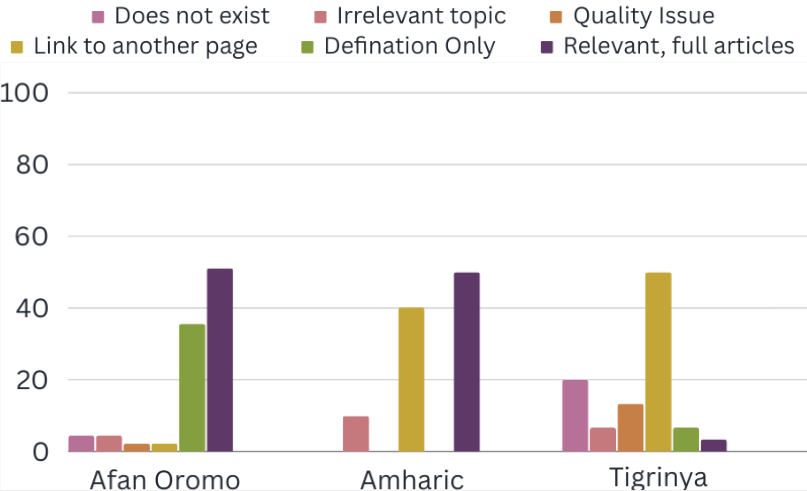}
          

          \caption{\textbf{Bar plots showing the different types of articles links on the main pages lead to for each of the three Wikipedia.} In this figure, we observe for Afan Oromo, around 35\% of the links in the main page link to one paragraph definition of the topic without further links. Additionally, we found 4.44\% of the links to lead to political posts that have nothing to do with the topic at hand and another 4.44\% to lead to non-existent pages. For Amharic Wiki, we found that 40\% of the links link to another category while
          9.9\% of the articles talk about COVID-19, unrelated to the category at hand, and in one case be an actual research paper published on Wikipedia. Tigrinya Wikipedia had the lowest percentage of articles leading to a relevant, full article (3.33\%). A majority of the links on the Tigrinya Wiki main page were links to other categories. Moreover, 13.33\% of the articles were stubs with just single words in another language or just one-sentence definitions of something related to the topic. Lastly, 20\% of the links led to pages that do not exist.}
      \label{fig:links}

        \end{figure}

We can also use main page links as a proxy for quality.  The Wikipedia main page for a given language typically links to various categories of articles---e.g., ``History.''  Fig. \ref{fig:links} shows the percentage of links in each of the three Wikipedias' main pages that lead to non-existent pages, irrelevant or off-topic pages, pages with quality issues, pages with just stubs, and pages with full articles and further links relevant to the topic. 
We also found links to pages that violate Wikipedia's policy~\cite{noauthor_wikipediareliable_2023} against opinionated political articles, harmful and abusive content, and full texts of research publications.

\section{Study 1: Analysing Forum Discussion by Wikipedia Contributors} \label{forum_analysis}

In Study 1, we collected and analyzed forum data (discussions among Wikipedia contributors) from Talk Pages \cite{talk_pages} on Wikipedia in Tigrinya, Amharic, and Afan Oromo. This study focused on understanding (1) issues raised by contributors on Wikipedia Talk Pages in the three low-resourced languages and (2) ways they navigate creating content in the face of those challenges.  

\subsection{Data Collection and Analysis}

\label{sec:collection}
We collected all the forum data available, with no exclusions, from all three Wikipedias in April 2023. Table \ref{tab:fourm_data} summarizes our dataset. We conducted an inductive thematic analysis~\cite{braun_using_2006} on the posts in Amharic, Tigrinya, and Afan Oromo. As an L1 Amharic and L3 Tigrinya speaker\footnote{L1=native language, L3=2nd non-native language, limited conversational}, the first author directly coded posts written in these two languages. The first author also provided translations for the posts that were in Tigrinya or Amharic. An external collaborator who is an L1 Afan Oromo speaker translated the posts in Afan Oromo before the first author coded the translations. The first author open-coded all posts line-by-line. Then, the first and third authors identified first-level themes from the open codes in weekly meetings over 2 months. Lastly, all authors met to refine the first-level themes and categorized them into second-level themes in weekly meetings over the course of 3 months. The entire process resulted in 265 unique open codes, 24 first-level themes, and 6 second-level themes.  Table \ref{tab:codes} in Appendix \ref{apd:Coding} shows our first- and second-level themes, with example open codes for each.

 In the rest of this paper, we refer to Wikipedia forum participants as ``posters.'' Poster details appear in Table \ref{tab:posters}. We identify each poster the following way: ``WP\emph{l}\emph{n}'' where \emph{l} is the first letter of their Wikipedia's language (A-Amharic, T-Tigrinya, O-Afan Oromo) and \emph{n} is 0 if they made the most edits among the posters in their Wikipedia and \emph{n} is 1 if they made the second most edits, and so on.
 For instance, WPO2 is a Wikipedia poster in Afan Oromo Wikipedia with the third most Wikipedia edits. This study was approved by our Institutional Review Board (IRB).

\begin{table}[]
    \centering
    \small
    \begin{tabular}{|c|p{1.5cm}|p{1.5cm}|p{2cm}|}
    \hline
         \textbf{Language} & \textbf{Number of Threads} & \textbf{Number of Posters} & \textbf{Time Span of Forum Data}\\
    \hline
         Amharic & 70 & 29 & 2005-2022\\
         Tigrinya & 10 & 14 & 2005-2019\\
         Afan Oromo & 6  & 7 & 2009-2020\\
         
    \hline
    \end{tabular}
    \caption{\textbf{Data collected from the Wikipedia forums for the three languages centered in this study.} In the forum, posters create discussion threads, each on a particular topic, in which posters deliberate over issues related to the topic.}
    \label{tab:fourm_data}
\end{table}

\begin{table*}[ht!]
    \centering
    \begin{tabular}{|p{1cm}|p{1cm}|p{1cm}|p{1.2cm}|c||p{1cm}|p{1cm}|p{1cm}|p{1.2cm}|c|}
    \hline
        \textbf{Poster ID} &\textbf{Total edits} &  \textbf{Local Wiki Edits}&\textbf{Account Created} & \textbf{Last Edit} & \textbf{Poster ID} & \textbf{Total edits} &  \textbf{Edits in Local Wiki} &  \textbf{Account  Created} & \textbf{Last Edit}\\
        \hline

        WPA0 & 40,625 & 20,425 & 2014$\oplus$ & 2019 & WPA18$\ast$ &  10,978 & 0 & 2005 & 2006 \\
        WPA1 & 13,754 & 13,558 & 2010 & 2020 &   WPA19$\ast$ & 7,395 & 0 & 2006 & 2019\\  
        
        WPA2 & 8,967 &  6,498& 2014$\oplus$ & 2018 & WPA20$\ast$ & 4,451 & 0 & 2007 & 2022    \\          
        WPA3 & 3,162 & 3,107 & 2008 & 2016 &  WPA21$\ast$ & 36,146 & 0 & 2004 & 2010  \\    \cline{6-10}                    
        WPA4 & 2,517 & 2,517 & 2009 & 2011 &  WPT0 & 375 &  324 & 2019 & 2022 \\ 
        WPA5 & 288 & 288 & 2006 & 2007 &  WPT1 & 39 & 39 & 2007 & 2007    \\

        WPA6 & 791,146 & 251& 2008 & 2020 & WPT2 & 121,453 &  7 & 2006 & 2006 \\  
        WPA7 & 598 & 180 & 2009 & 2010  & WPT3  & 33,384 & 2 & 2009 & 2009  \\   
        WPA8 & 1,175 & 64 & 2006 & 2016 & WPT4 & 122,481 & 2 & 2008 & 2010 \\ 
        WPA9 & 66 & 64 & 2010 & 2011   & WPT5$\pm$ & - & 1 & 2008 & 2008 \\  
        WPA10 & 41,461 & 22 & 2006 & 2008   &  WPT6 & 11,056 & 1 & 2008 & 2008  \\
    
        WPA11 & 129,170 &  16 & 2008 & 2012 & WPT7$\pm$ & - & 1 & 2015 & 2015   \\ \cline{6-10}
        WPA12$\pm$ & - & 15 & 2010 & 2010 & WPO0 & 29,987 & 1771 & 2020 & 2022  \\
        WPA13 & 151,263 & 10 & 2010 & 2016 &  WPO1 & 1,026 & 630 & 2020 & 2021 \\
        WPA14 & 1,911 & 7 & 2021 & 2022 &  WPO2 & 19 & 16 & 2020 & 2020 \\
        WPA15 & 6 & 6 & 2020 & 2020 &  WPO3 & 791,151 & 12 & 2010 & 2013\\
        WPA16 & 33,384 & 2& 2009 & 2009 &  WPO4 & 33,384 &  2 & 2009 & 2009\\
        WPA17 & 1 &1 & 2015 & 2015 & &  & & \\
\hline
    \end{tabular}
    \caption{\textbf{Table showing details of forum participants.} We arranged the table by decreasing order of contribution size to local Wikipedia and arranged the languages per decreasing size of posters (Amharic 22 posters; Tigrinya 8 posters; Oromo 5 posters). Of the total posters per forum 7/29 for Amharic, 6/14 for Tigrinya, and 2/7 for Afan Oromo no longer have active accounts or did not sign their posts so we could not get their details. We indicate with ($\ast$) where posters are from a different Wikipedia (English, Kurdish, Nepali) but posted in the forum. We used the Global User Contributions tool~\cite{global_2023} to get the number of total edits and local wiki edits. Accounts with a higher number of total edits compared to local edits have associated accounts in multiple languages' Wikipedia and contribute to Wiki sources like Wiktionary. With ($\pm$), we indicate posters that have IP addresses displayed in place of usernames; for those accounts, we used the earliest year of editing as the `Account Created' year. Additionally, we use ($\oplus$) sign in the Account Created column for posters that do not have their account creation date listed on their accounts; for those posters, we used the earliest year of edit as a substitute. }
    \label{tab:posters}
\end{table*}


        

    

\subsection{Findings} \label{forum_findings}
In this section, we discuss the themes that emerged from our analysis, with each sub-section relating to a second-level theme. The forum discussions include Amharic, Tigrinya, Afan Oromo, and English text.  

\subsubsection{Posters use idioms, statistics, and role divisions to keep their Wikipedia growing.}
We found that posters interacted to discuss strategies on how to grow their Wikipedia. One avenue discussed by posters (WPA7, WPA15) is to publicize the Wikipedia on professional networks like the ``Ethiopian Chemists Association'' and ``Engineers Associations'' to get more people involved in writing. Another sub-theme that emerged is using the number of articles created in different time spans to motivate contributors. Some posters also expressed frustration over the growth rate and used article numbers from other languages to compare the growth of their Wikipedia. There were also posts using idioms and congratulatory phrases to motivate the creation of more articles. WPO0 commented ``I also really like the work being done on this site [Wikipedia] in this [Afan Oromo] language.'' In our analysis, we also noticed that contributions are not limited to native speakers. Non-native speakers supported the growth of the Wikipedia by bringing articles for translation, or by writing in the local language and requesting editing support from other contributors.

\subsubsection{Latin-based keyboards, platform updates, and heavy fonts make writing and reading in non-Latin scripts difficult.} \label{forum_non_latin}

From our analysis, we found that 5 out of 29 posters in Amharic Wikipedia discuss the difficulty of writing in Ge'ez due to (1)~the Wikipedia interface not supporting input in the script, especially after an update, and (2)~challenges with using a Latin-based keyboard to write in a non-Latin script. As WPA2 commented ``a bigger problem is that currently, you can not type Amharic using the new interface.'' In a separate thread, WPA0 and WPA1 discussed the challenge of typing in Ge'ez script on a Latin keyboard, pointing out that it is a time-consuming process. Challenges with the Wikipedia interface in non-Latin scripts do not stop at inputting text; 4 out of 14 posters in Tigrinya and 7 out of 29 posters in Amharic Wikipedia also had difficulties with the fonts for Ge'ez. WPA1 and WPA2 complained that the ``default [Wikipedia's Ge'ez] font'' looked ``very distracting'' and asked for alternative fonts to be added. The font designer, who does not read Amharic, misunderstood the issue and responded that the font does not have a bold variant. Because the designer themself did not read Amharic, it took several rounds of back and forth before they understood that the issue was that the font looked ``bold'' for all text (in other words, the width of the strokes was too heavy).

\subsubsection{Entity extraction, new articles, and access to media and funds are among the common asks on Wikipedia forums.} \label{forum_requests}
 
   We observed multiple threads of discussion in the forum data centered around finding the translation of entity names from a given set of English, Amharic, and Tigrinya documents. WPO3 on the Afan Oromo forum requested confirmation for an entity name they translated, while another poster corrected their spelling. WPA6 requested the creation of 16 new articles from given topics, in some instances providing the English version and links to media. An interesting interaction we observed is that requests are not limited to the Wikipedia of the given language. For instance, in the Amharic Wikipedia forum, WPA6 requested help finding entity names in Tigrinya from a given document. The discussion threads showed back and forth between WPA6, WPA2, and WPA0 around what the translations are, where the entities are located in the documents, and which formats made the most sense to native speakers. Another set of requests we observed was in ways of collecting supporting media, such as pictures, to support their articles. WPA6 indicated they were not located within Ethiopia and requested those who were located there to take pictures of certain places (e.g. Airports) for them. WPA1 posted about potentially traveling to Ethiopia to collect media for their articles. Wikipedia posters also discuss the funding source of Wikipedia and ask who has access to the funds. One poster, whose account is no longer active, also asked for support in terms of mobile data top-up in discussing how to increase article contribution. WPA7 indicated there has to be a budget to access resources that would help Wikipedia in Amharic grow.

 \subsubsection{Wikipedia posters from different languages interact on forums to collectively grow their individual Wikipedia.}
In the Amharic Wikipedia forum, we found a discussion thread about potentially creating a Wikipedia page for Ge'ez under the Amharic Wikipedia. This discussion involved a back and forth about creating an independent Wiki for Ge'ez and ended with how Wikipedia for Afar (a Cushtic language spoken in Ethiopia) was under threat of closure by Wikipedia since it did not have enough contributors. Currently, the Afar Wikipedia is not active. Beyond the local languages, contributors to other languages' Wikipedias also post in the forums requesting actions related to their languages, such as: (1) to add articles that are translations of articles from Polish Wikipedia, (2) to add a link to the Amharic Wikipedia into a list of languages on Nepali Wikipedia, (3) to add a link to the Tatar language Wikipedia on the Tigrinya Wikipedia front page, or (4) to translate an Amharic document to English, in support of an English Wikipedia article.

\subsubsection{Human and machine translation are among the suggested alternatives for growing low-resourced language Wikipedia.} \label{forum_translation}
 
 Some posters suggested translation as a way to mitigate the challenges of typing in non-Latin scripts (Section \ref{forum_non_latin}). 5 out of 29 posters in Amharic and 2 posters in Tigrinya Wikipedia also discussed issues around how interface words are in English. In the Tigrinya Wikipedia forum, WPT7 provided translations for common interface words. In the Amharic Wikipedia forum, there was discussion around using translatewiki \cite{BibEntry2023Dec} and translation memory for commonly used interface words and system messages. WPA6 also took the role of bringing articles from English Wikipedia that are about topics related to Ethiopia and requested translations from other contributors. Two posts in Amharic Wikipedia were also dedicated to finding Tigrinya and Afan Oromo speakers who could translate an article that was created in Amharic into the two languages.

%

\subsubsection{Misspellings and language mixing lead to duplicate articles, affect search experience and consume editing time.} \label{misspellings}

 Posters discussed how a lack of standardized spellings for words and phrases leads to duplicate articles about the same concept. One challenge mentioned in the Amharic Wikipedia forum is that of words with the same pronunciation and meaning but different spellings (see Section \ref{background}) and their impact on the user's search experience. One poster gave a detailed explanation supported by examples of their own search experience, elaborating how there would be ``182 ways of writing `Aste Hailesielase'.\footnote{Ethiopian King from 1930-1974 Gregorian Calendar.}'' We saw mixed sentiments on this issue: posters like WPA15 called on Wikipedia to handle this issue and asked that their search experience accommodate this characteristic of the Amharic language while others asked for the normalization of words and a way to equate all of these words. One poster opposed the proposal for normalization, indicating it would lead to the loss of the characteristics of the language. In Afan Oromo Wikipedia, posters WPO1 and WPO2 had concerns over widespread misspellings in articles and that it ``may render the Oromo section of Wikipedia unusable.'' Posters indicated this issue trickled down to their search experience: WPO2 elaborates, ``You cannot search and find topics [on Wikipedia] b/c they are misspelled when the topic is created.'' WPO1 confirms that misspellings have led to topic duplication: ``There are topics I created but later noticed it exist[s] with different/wrong spelling.'' Another quality issue was English-mixed content: in the Tigrinya Wikipedia forum, there are complaints of articles being in English, which puts an additional burden on the posters to identify and edit the articles.

\section{ Study 2: Contextual Inquiry Study with Novice Contributors} \label{contextual_inquery}

\hyperref[forum_analysis]{Study 1} allowed us to explore challenges faced by contributors who are already experienced and engaged with Wikipedia enough that they post on the Wikipedia forums. In Study 2, we were interested in observing the challenges that obstruct \textit{novice} contributors as they start attempting to contribute to Wikipedia. We conducted a contextual inquiry \cite{holtzblatt_3_2017} study followed by semi-structured interviews; for 14 participants, we first observed the participant writing an article in one of the three low-resourced languages, then asked follow-up questions to improve our understanding of their process.  For this study, our participants were Wikipedia novices, with 10 out of 14 never having contributed to Wikipedia.

\begin{table*}[ht!]
    \centering
    \small
    \begin{tabular}{c|p{1cm}|p{2.8cm}|p{2.3cm}|c|p{0.8cm}|p{1.3cm}|p{0.9cm}|p{1.1cm}}
    \toprule
        \textbf{ID} & \textbf{Age Range} & \textbf{Field}& \textbf{Profession} & \textbf{Language} & \textbf{Device}  & \textbf{Location} & \textbf{No. Articles}& \textbf{Telegram Channel}\\
    \midrule
        P0 & above 40 & Geo-informatics&Lecturer & Tigrinya & Laptop   & Ethiopia & 0 & No\\
        P1 & 26-30 & Philosophy&Grad Student & Afan Oromo & Phone  & Ethiopia & 6  & Yes\\
        P2 & 18-25 & - &High School Student & Amharic & Laptop   & Ethiopia & 1 & No \\
        P3 & 26-30 & History&Professor & Amharic & Laptop  & Germany & 5 & Yes\\
        P4 & above 40 & Language&Lecturer & Tigrinya & Laptop  & Ethiopia & 0 & No\\
        P5 & 30-40 & Computer Science&PhD Student & Tigrinya & Laptop  & Germany & 0 & No\\
        P6 & 18-25 & Software Engineering& BSc Student & Afan Oromo & Laptop  & Ethiopia & 0 & No\\
        P7 & 18-25 & Geology& Packaging company employee  & Afan Oromo &  Phone  & Ethiopia & 0 & Yes\\
        P8 & 18-25 & Electrical Engineering &BSc Student & Amharic &  Phone & Ethiopia & 0 & No\\
        P9 & above 40 & Language&Teacher & Tigrinya & Laptop  & Ethiopia & 0 & Yes\\
        P10 & above 40 & Social Sciences &NGO employee & Amharic & Laptop & Ethiopia & 2 & No\\
        P11 & 26-30 & Journalism& Artist, Writer & Amharic & Laptop & Ethiopia & 0 & Yes\\
        P12 & 30-40 & Computer Science& Engineer & Amharic & Laptop & Ethiopia & 0 & No\\
        P13 & 18-25 & Biomedical Engineering& MS Student & Afan Oromo & Laptop & Ethiopia & 0 & No\\

    \bottomrule
    
    \end{tabular}
    \caption{\textbf{Table showing participant details of contextual inquiry study.} We interviewed 14 participants and observed as they wrote an article on Wikipedia in one of the three languages that are the focus of our study. Our participants had diverse professional backgrounds and came from different age groups. Additionally, they used different devices when writing on Wikipedia; 3 participants used their mobile phones while the remaining 11 used Laptops. in No. Articles column, we give the number of articles our participants reported they had contributed to Wikipedia in the past. The Telegram channel column indicates whether or not the participant owns or administers a Telegram channel where they create and post content in a local language.}
    \label{tab:particpants}
\end{table*}

\subsection{Methodology}

\paragraph{Participants}
 We recruited participants via social media (Telegram), Wikipedia Forums, and using the authors' networks. We primarily targeted Telegram channel contributors that are known to the first author, as Telegram is popular in Ethiopia \cite{noauthor_chapa_2023}. We used a screening survey to select participants from several domains. Our aim with this study was to observe a wide range of user challenges, in all three languages and reveal the needs of novice low-resourced language speakers who interact with OKRs. As such, our main selection criteria was the ability to read and write in one of the low-resourced languages. We received 43 responses to our survey and contacted 22 respondents prioritizing language diversity, exposure to Wikipedia, and experience writing content in local languages on other platforms. Of the 22 respondents, 2 did not respond and 6 could not find a suitable time to meet with the researcher. Table \ref{tab:particpants} provides information about our participants. The first author's community membership was useful in identifying community practices for communication, such as using Telegram as a primary form of communication with participants instead of email, identifying the proper way to compensate participants located in Ethiopia, and interchanging languages to adapt to participants' comfort.



\paragraph{Consent and Compensation}
This study was approved by our IRB. Before participating in the study, participants signed a consent form per our IRB protocol. We also asked each participant for verbal consent before recording the session. Participants received compensation at a rate of 25 USD Amazon gift card or a 1000 ETB mobile top-up per hour of participation, depending on their location.

\paragraph{Session Structure}
The first author served as the interviewer and transcriber. Each study session lasted 48-130 minutes, with session times varying according to how long it took a participant to author their Wikipedia article. Each session included an initial observation phase followed by a semi-structured interview. We conducted sessions remotely over Zoom or Meet and recorded them for subsequent analysis. During observation, we asked participants to share their screen and think aloud \cite{book_think_out_loud} as they wrote an article in a language and a topic of their choice. During the semi-structured interviews, we raised our observations with our participants to confirm or refine our interpretations of their actions. One interview was primarily conducted in English with some mix of Tigrinya; all other interviews were primarily conducted in Amharic with a mix of Tigrinya and English, depending on participant preference.

\paragraph{Data Analysis}
 We conducted inductive thematic analysis \cite{braun_using_2006} of 20 hours of footage from 14 sessions as video recordings of the observations and semi-structured interviews using QualCoder \cite{qualcoder_2023}. The first author, as the only author able to speak Amharic or Tigrinya, associated short, descriptive sentences of participants' behaviors with segments of the video recordings in the open-coding phase. Then, in subsequent weekly meetings, the first and third authors discussed and refined open codes and synthesized first-level themes. The first and third authors then deliberated over the first-level themes, grouping them into second- and third-level themes and arriving at three hierarchical levels. The final hierarchy of themes was discussed between the first, second, and third authors and refined accordingly in weekly meetings. The analysis took 5 months to complete. Our analysis led to 443 unique open codes, grouped into 36 first-level themes, 9 second-level themes, and 4 third-level themes. We present the higher-level themes and example codes in Table \ref{tab:study_2_codes} in Appendix \ref{apd:Coding}.

\subsection{Findings} \label{contextual_findings}
In this section, we will discuss the themes that emerged from our analysis, with each subsection relating to the 4 third-level themes described in Appendix \ref{apd:Coding}. In Figure \ref{fig:timeline}, we provide the timeline for P9, showing most of the challenges we discuss below. 

\subsubsection{Lack of Language Support Tools} \label{lang_support}
We observed that language technologies, such as keyboards and translation tools, present their own set of challenges for low-resourced language speakers interacting with OKRs. We observed that some of our participants (P1, P7, P8) used their phones while all others used their laptops. Some participants (P8, P11) indicated using their phones was ``easier since the keyboard is visual'' and that they prefer writing on their phones and copy-pasting to other places. 

\begin{quote}
    I usually prefer writing with a [regular] pen\ldots{}the [key]board is not comfortable. Unless I write in Latin but that is not nice. It [the current keyboard] breaks your thought flow when you have to make so many decisions\ldots{}it takes time to find letters. We have over 26 or 27 letters each with 7 versions\ldots{}finding the letters is hard. --P8
\end{quote}

Some of our participants (P4, P9, P11, P12) could not get their keyboards to work on Wikipedia.  All of these participants used a Latin-based physical keyboard paired with software that transformed keypresses to allow them to write in Ge'ez. 
These participants ended up (1) restarting their computers (P9), (2) writing in other software like Microsoft Word, then copying and pasting (P2, P4, P9, P11), or (3) changing their browser from Microsoft Edge to Opera (P12). For P2, editing outside of Wikipedia resulted in an `Invalid Session' notice when he came back to publish the text. 
We also observed that our participants used seven different types of digital keyboards.  These included (i)~software that transformed keypresses from a Latin-based keyboard~\cite{keyman_2023, power_geez, vis_geez, geez_ime_windows, BibEntry2023Dec} and (ii)~software keyboards for use with touchscreens---e.g., for use on mobile phones~\cite{fyngeez_2023,geez_ime_android}. Most of our participants' keyboards only supported changing the character of the consonant when a vowel is written next to a consonant  (see Section \ref{background}). Hence, when participants wanted to write a word that had a consonant followed by an independent vowel, we observed participants (P0, P4, P5, P8, P9, P12) writing a space after the consonant, writing the vowel, and then removing the space; a process which resulted in spelling errors for some participants (P5, P9). Another issue with current digital keyboard setups is in special characters. Special characters, including some punctuation marks, are reserved as vowel markers or as different letters in the Ge'ez alphabet. Hence, when participants needed to use those characters they had to either turn off their current keyboard (P11) or go to symbols (P0, P9).  We also observed participants (P0, P1, P9) having issues with letters that do not have direct Latin equivalents, thereby needing multiple keystrokes to write a single letter: 

\begin{quote}
    For instance to write one letter [k'\textsuperscript w\"a] I have to use [keys for] Q plus W plus A\ldots{}three times\ldots{}three different actions needed. You see how long it takes me to write just one paragraph? Imagine how much longer it is going to take to write a full article. --P0
\end{quote}

We also observed participants (digitally) switch keyboards in the middle of writing. P5 had to switch keyboards to enter his email while signing up. P8 switched his keyboard to write the word ``muse'' in parenthesis since he did not have a perfect Amharic equivalent for it. Once he wrote the word, he forgot that he had switched keyboards and continued writing Romanized Amharic (using Latin characters instead of Ge'ez) until he realized what he was doing and rewrote it using Ge'ez.

Participants also raised questions about the modality of input for knowledge repositories. P7 explains how she cannot find online any ``Walaloo'' (poems) and stories her father used to tell her and her siblings as kids. P7 further elaborates ``He wrote them in a big book but we got them through listening to his stories at night. But we are losing that culture now.'' Another participant (P11) explains how the stories that are lost are usually from minoritized groups within the society:

\begin{quote}
    When I was doing summer work in Southern Ethiopia, I asked my students to bring stories about a prominent figure in their community. One student brought an article about a powerful woman called Aku Manoye from Sidamo, a woman with similar characteristics as Kake Wordot \cite{yekake_wurdowet} from Gurage and many others from Arsi, Gedio e.t.c. I never heard about Aku Manoye before\ldots{}so I was really surprised. I believe that a lot of stories like this, especially the ones about women, are buried in oral traditions we do not get to experience. Writing is confined to a handful of languages and even within those languages, there is further marginalization of women. --P11
\end{quote}


Our participants (P0, P11, P13) indicated issues with translations of scientific terms, saying the phrases are ``\textit{usually hard for everyday use}'' and that for Tigrinya and Amharic, they are ``taken to the Ge'ez roots'', which are not used in everyday conversations. We observed a similar concern for translating scientific terms to Afan Oromo. As P13 notes, ``the translations are not exact but\ldots{}it hints at it. For example, AI in Amharic is `Sew Serash Astewelot' \ldots{}in Afan Oromo it is `hubannoo nam-tolchee' \ldots{}same direct translation.'' P0 further discusses the translation for Artificial Intelligence to Tigrinya: 

    \begin{quote}
        It is hard to translate scientific terms. I was in a conversation with a linguist recently on translating Artificial Intelligence to Tigrinya. The suggested translation was `Seb Serah Lebwo' or `se. se. le.' for short. It is related to the Ge'ez root word of the concept of intelligence and artificialism. Further, the linguist suggested we abbreviate the term, similar to how Artificial Intelligence is referred to as AI. The translation makes sense in retrospect but we do not really use Ge'ez words in everyday communication. It is new for all of us and it is hard. --P0
    \end{quote}

    In addition to translating scientific words, we observed our participants used translation (both human and machine) as they were writing their articles. Participants used translation to find words or spellings they forgot (P1, P6, P7, P10), to paraphrase an English article into a local language (P6, P5), or to translate portions of English text directly into a local language (P2). However, there were concerns with using translation, whether machine or human; the main concern was the loss of meaning, especially in concepts and words that do not exist in one language. 
    
    One issue we observed with Machine Translation (MT) systems was the lack of cultural contexts such as historical events that led to the loss of meaning. We observed that translation systems would sometimes transliterate the words instead of translating them (P2, P12). P2 notes how the MT system did not translate the phrase ``Christian highlands'' but just put the transliteration in Ge'ez characters: ``But `highland' is bottled water in Ethiopia.'' \footnote{Highland was the name of the first bottled water company in Ethiopia and the term has since been adopted to refer to bottled water in general when used in Amharic.} Participants (P1, P2, P5, P6, P8, P10, P11) also criticize current machine translation systems for being ``\textit{too dry, too literal and direct}'' and for ``\textit{lacking context.}'' 
    \begin{quote}
    It is getting better but sometimes it messes\ldots{}like everyday things. It is very direct. Like if I try\ldots{}`ehel wehachen aleke.' it returns [waits for the response] `we ran out of grain and water.'\ldots{}very direct translation. But an average Amharic speaker can tell you `ehel wehachen aleke.' means our time[usually romantic] has come to an end. I also noticed it makes more mistakes when the sentences are long. If I want it to translate well, I use short sentences --P10
    \end{quote}

    We also observed translation outputs that had unrelated political outputs (P2, P6) or returned alarming outputs (P2, P6). When P2 was translating an article about a famous Ethiopian Muslim ruler, the output had a sentence that said in Amharic `He married his daughter', indicating the ruler married his own daughter. However, the original text had a third person, whose daughter was wed to this ruler. The translation missed this context entirely. Our participants (P2, P6, P7, P10) also noted gender bias in machine translation outputs, with outputs leaning towards the male gender, even in cases where gender-neutral options are available. As P6 was translating an article about Sinqee, a women's organization in the Oromo culture, she noted a male gender bias in the output: ``it says `abbaa [father] sinqee.' We do not have that\ldots{}We have `haadha [mother] sinqee'.'' Additionally, Afan Oromo translation outputs also included a mix of English (P6, P7). After trying to translate an English article to Afan Oromo, P6 observed the following:

    \begin{quote}
        First, we do not say `Dhaabbata' to refer to these things\ldots{}it is directly translating `institution'\ldots{} Second, this one translates to `death institution.'\ldots{}The institution of homicide\ldots{}It says `killing or murder institution.' There are better translations in Afan Oromo. The article talks about Gumaa, it is a conflict resolution system used when a homicide happens\ldots{}it is not a system for murder or a killing institution. If this was used in [the] news, it might create another issue.'' --P6
    \end{quote}

    Another issue we observed with current translation mechanisms is the lack of support for a different calendar system (Section \ref{background}). Participants manually did the conversion (P10, P13), dropped the date and only retained the month and year (P2, P12), or added a marker to indicate the calendar in use (P13). Overall, we observed using translation required a lot of post- and pre-editing (P2, P6, P10), had errors that significantly changed the meaning of the article (P2, P7, P6), and resulted in inconsistent date formats (P2, P10, P12).

\subsubsection{Socio-Political Issues} \label{socio_findings}
The challenges we observed in our study are not purely technological barriers but are at the intersection of social, political, and technological aspects of communities. We observed several participants (P2, P7, P9, P11, P12) got a ``Permission Denied'' response from Wikipedia servers because they were connected to Virtual Private Networks (VPNs) and tunneling their traffic to circumvent the Internet Providing Service's blocking of certain platforms like Telegram \cite{ethiopia_internet_shutdown}. All five participants tried turning off their VPN, and only one participant (P7) was able to continue with the session after turning off their VPN while all others got disconnected from the internet. Once they reconnected, P12 used a different browser to try to create an account with the VPN turned off. P9 tried using the same browser but still got a ``Permission Denied'' response from Wikipedia when attempting to edit articles. Finally, P9 used Microsoft Word to write his article but was not able to copy it over to Wikipedia. P9 said ``I am back after a year. Here is the problem: first you have to open a VPN, then you connect to the VPN service and access certain platforms. So much setup before you can get to where you were first.''

Another issue that is at the intersection of technical and societal challenges is the lack of scholarly work for participants (P1, P5, P6, P7, P8, P13) to cite. For some participants (P0, P5, P8), the resources they have are locked in hard-copy books. Additionally, our participants (P2, P3, P5, P9, P10) note how most research publications about their communities are of European sources. ``See? it is the Harold's of the world.'' says P3 scrolling down the reference link in a Wikipedia article about one of the Ethiopian kings and looking at the authors of the cited sources. P3 further elaborates how, when she once tried to publish an article about a famous Ethiopian person on English Wikipedia, her article was rejected by the editors since she used sources that were written in Amharic and the ``reviewers did not understand Amharic.'' As P2 states, ``Most history is written from one side. There is a lack of representation in historians, even in local ones. Only bad things are written about this leader (referring to a Muslim Ethiopian ruler).'' In addition to accessing resources that are written in local languages and with local context, our participants (P0, P5) note how research conducted locally is written up in English. 

\begin{quote}
    The medium of instruction is in English but in classrooms, we translate to a local language, Amharic or Tigrinya. We talk in local languages but to publish or write, we use English. We do not exercise writing in these [local] languages. Even when we work on research, we first collect data questionnaires in Tigrinya or Amharic, then we translate the results to English to publish so it `becomes research.' It is so stunning and ironic! --P0
\end{quote}

We observed that participants had a difficult time finding resources in their languages and in their context both on popular search engines and on Wikipedia. Our participants (P1, P5, P6, P7, P8, P13) also pointed out that when they search for content in their local language on Google, they ``get content in other languages'', ``usually just get social media posts'' or ``Facebook links'', which they comment are ``usually political''.  One participant (P6) found political posts unrelated to the title in Afan Oromo Wikipedia. We also observed three of our participants (P2, P3, P11) indicating surprise at the sight of the Big Mac article in the front. ``It says Big Mac in the front\ldots{}hmm\ldots{}why is that here?'' The lack of contextually and linguistically relevant articles also trickles down to search experience: some of our participants (P2, P5, P12) had issues finding articles to edit. P2 wanted to edit biographic articles about Africans and adjusted the search filters to ``Biography'' and ``Africa''; he got article suggestions for Ukrainian president Volodymyr Zelenskyy instead. P5 had a similar experience trying to edit historical articles about Africa, after adjusting the filters, he got a notice saying ``No suggestions found'': 

\begin{quote}
    hmm\ldots{}history\ldots{}I want to write about history\ldots{}and then for what region? for Africa\ldots{}and then what do I do?[\ldots{}]tsh [disappointment]\ldots{} we have nothing\ldots{}see?\ldots{}we don't even have suggestions\ldots{}should I change it so we get more suggestions? For example, will we get suggestions if I say America or Europe? Maybe\ldots{} --P5
\end{quote}

\begin{figure*}[!tbp]
          \centering
          \subfloat[Participant first searched for the name of the town in Ge'ez, `Sazat', mixed with the English phrase ``sazat tigray''. The first search result as well as the 'frequently asked questions were all about the recent war in the region. There were no results for the town.]{\includegraphics[width=0.48\textwidth]{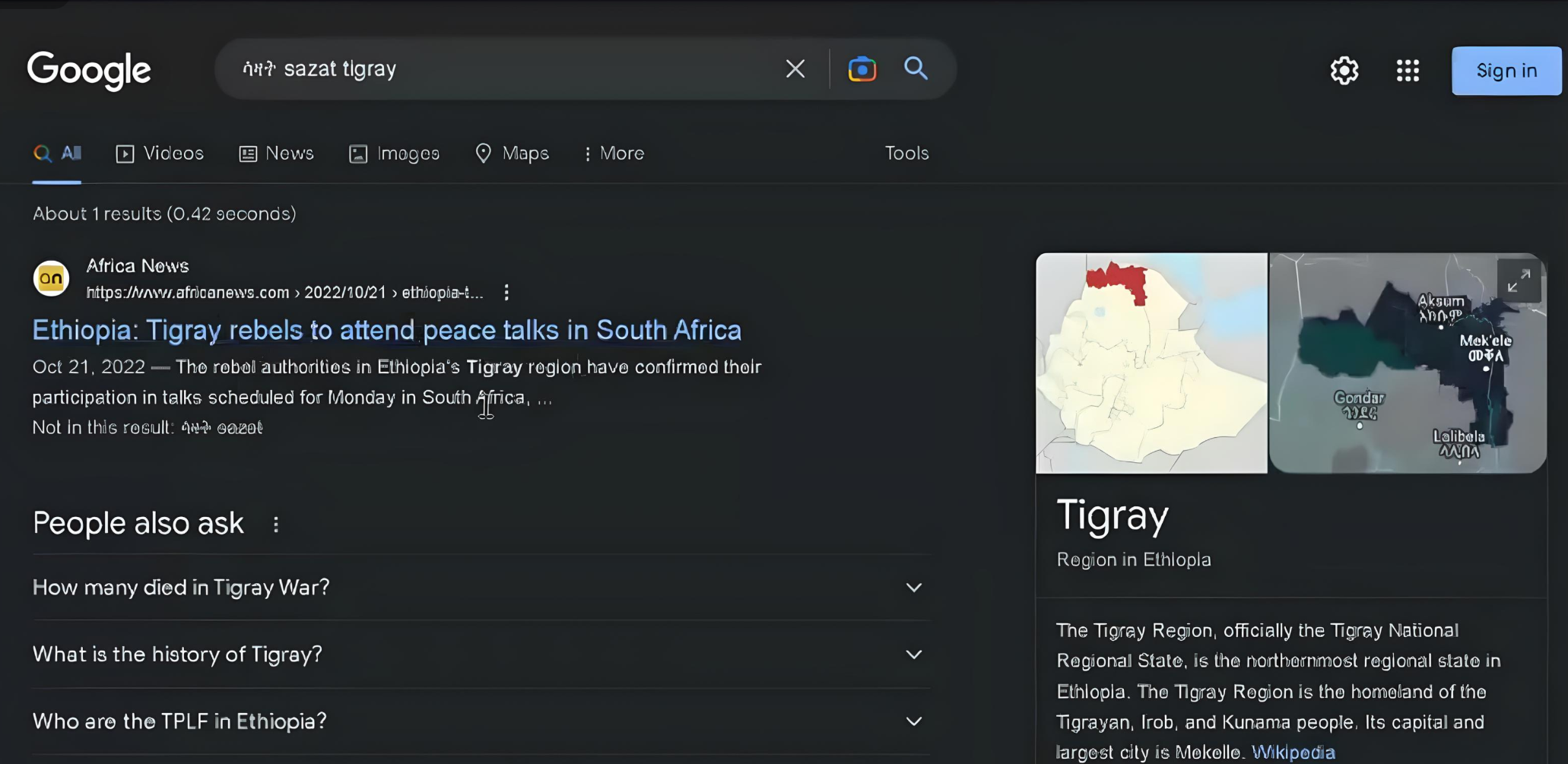}\label{fig:search_both}}
          \hfill
          \subfloat[The participant then searched for the name of the town in Ge'ez and removed the English description. In this case, they got videos from YouTube with phonetically similar but conceptually unrelated content with description: ``Sayat yemtmarkegn sazat yemtetazezegn'' meaning ``A woman who appeals to me when I see her and obeys my order'']{\includegraphics[width=0.48\textwidth]{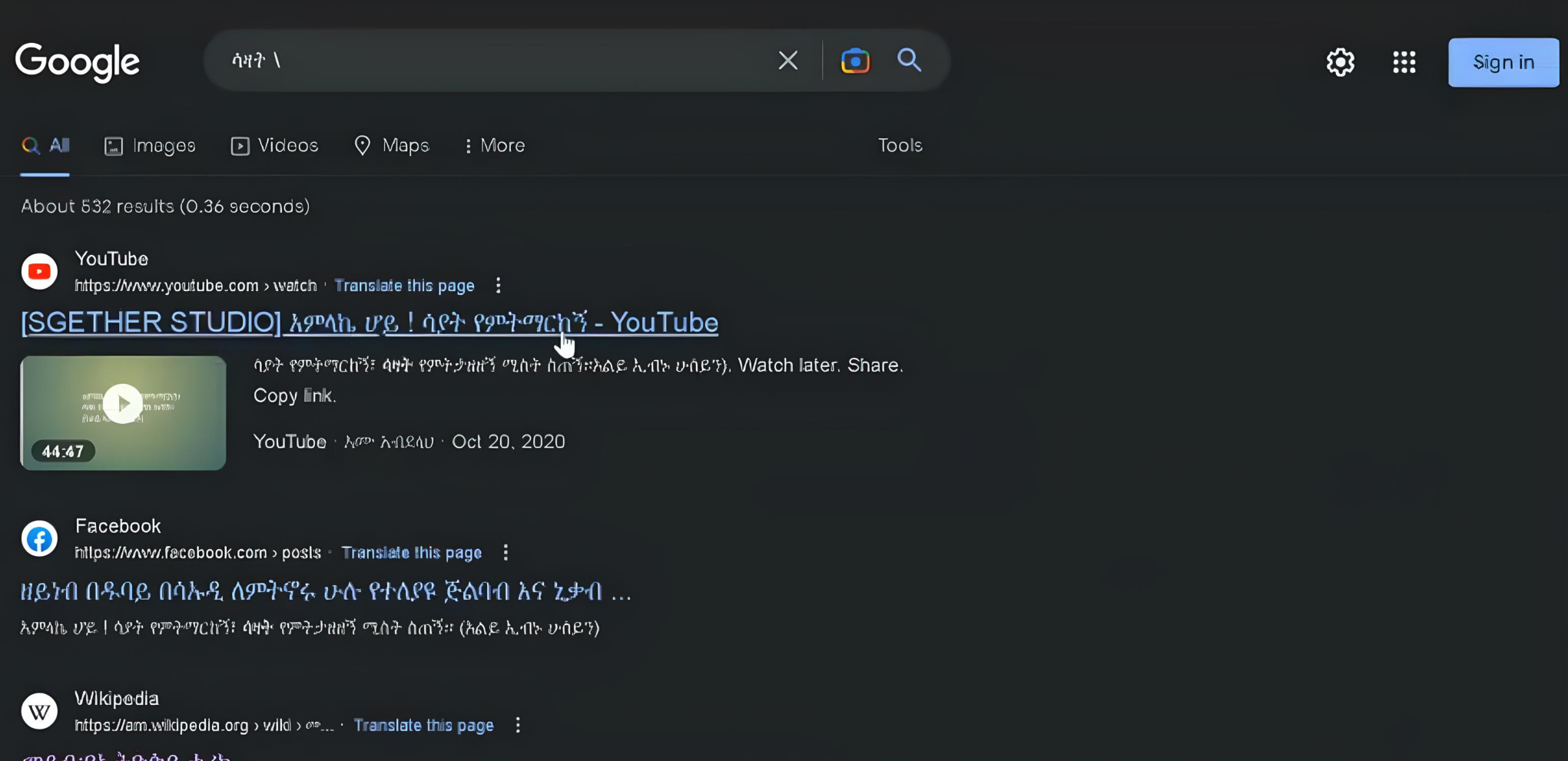}}
          \hfill
          \subfloat[The participant then searched for ``Sazat, Tigray Mesraq'' meaning ``Sazat Tigray East'', adding more description in Amharic. They got a page indicating no search results could match their query.]{\includegraphics[width=0.48\textwidth]{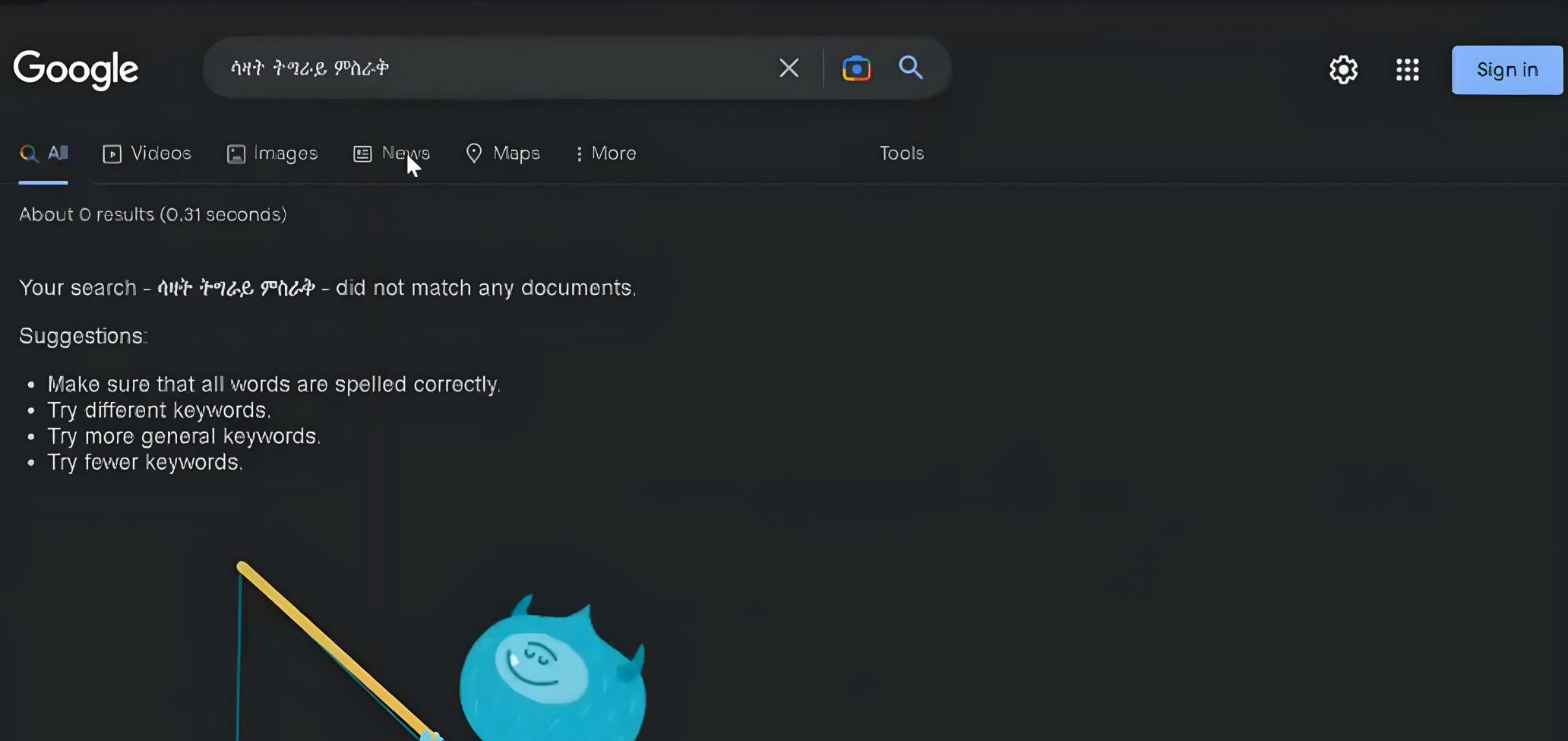}}
          \hfill 
          \subfloat[Finally, the participant copied a portion of text from the Wikipedia article they were editing and searched for it on Google, only to be told their search results did not return anything again.]{\includegraphics[width=0.48\textwidth]{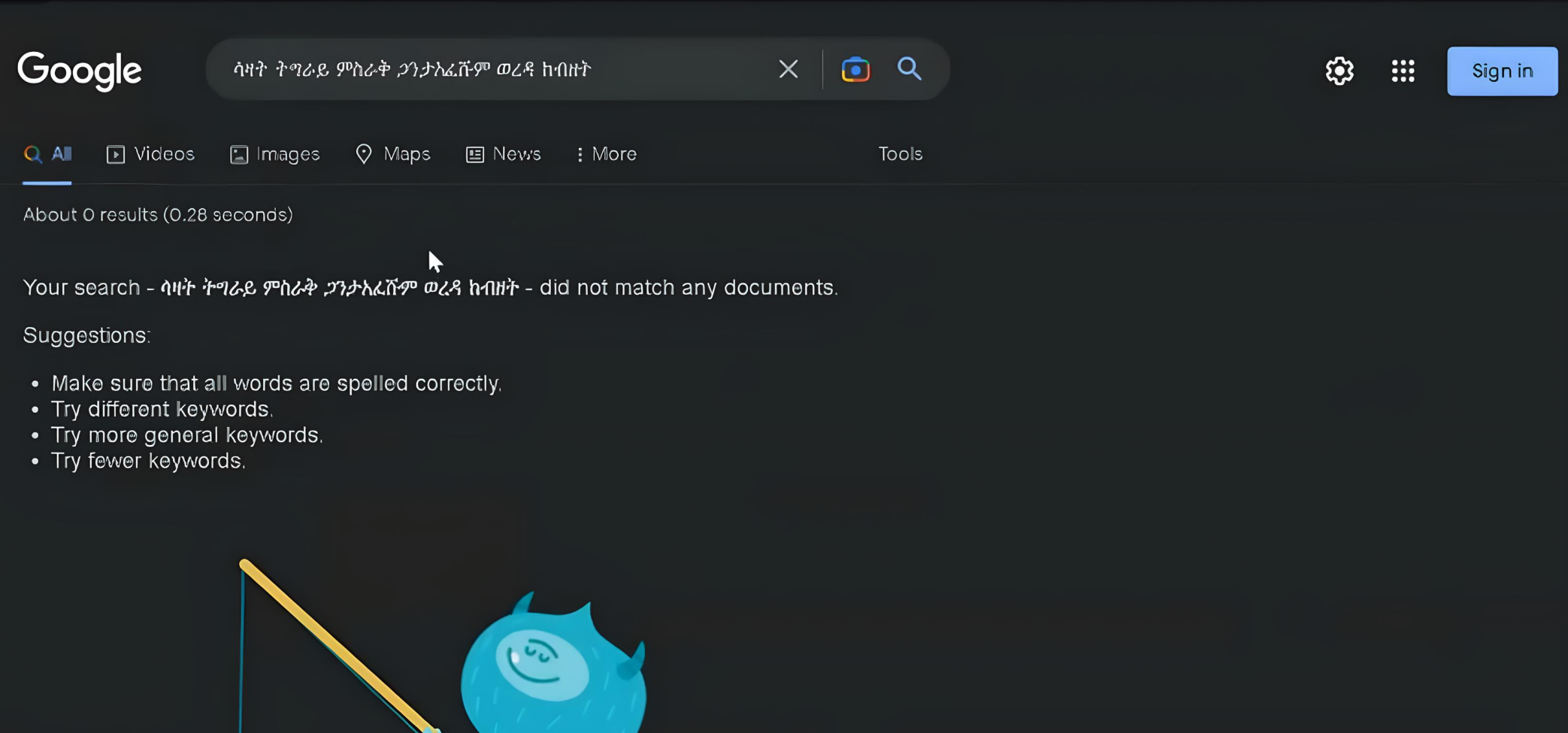}}
          \hfill

          \caption{\textbf{Figures showing screenshots of the interaction of P12 in trying to find references for an article they wanted to edit on Wikipedia about a town in the Tigray region of Ethiopia.} The search experience resulted in contextually irrelevant results and was skewed by recent events. Finally, the participant changed the topic they wanted to write about because they could not find resources. All search was conducted on google.com after the participant disconnected from VPN.}
      \label{fig:search}

        \end{figure*}

We observed that our participants have three ways of dealing with the lack of sources to cite in local languages: (1) find a source in English (P2, P11), (2) create the resources they needed (P2, P10), (3) use Wikipedia itself to find sources (P2, P3, P7), or (4) change the topic (P5, P12). As an example, P12 wanted to edit an Amharic Wikipedia stub about a town in Tigray and searched on Google for the name of the town followed by the phrase ``Tigray region.'' Fig. \ref{fig:search} shows the steps P12 took trying different search queries to find resources before he finally gave up.

 Our participants (P0, P4) also raised financial barriers to publications, limiting the amount of scholarly work contributors can reference when writing articles in their languages. P0 notes how there is a lack of affordable places to publish scientific papers, and how the problem becomes even harder if one tries to publish in a local language. As P4 notes in our interview, financial barriers also limit \emph{access} to existing publications: ``There are books that our university does not have access to because our university does not have an agreement. I can see the books [metadata] but not be able to access them.''
 
\subsubsection{Navigating Wikipedia's Interface Challenges} We observed that our participants had trouble with the Wikipedia interface due to unfamiliar interface words, instructions in mixed languages, and insufficient information on how to perform actions. We observed participants' (P0, P2, P3, P6, P7, P10, P11, P12) confusion over what different interface words meant, which in some cases resulted in participants (P1, P10, P11) being unable to find the right button or link. We found three major reasons for the confusion: (1) participants (P0, P2, P3, P11, P12) described the interface words were ``not common, every day words'', (2) participants talked about how they were ``used to these [interface] words in English as that is how we were introduced to technology.'', and (3) participants (P4, P5, P6, P7) noted how different dialects might be the reason they do not understand the interface words.

\begin{quote}
    I do not know what `Lakkaddaa' means but `gulaali' means `examine' so I guess this is edit? Dialect--in Afan Oromo we call it `loqoda'--it might be the reason. I think this is another dialect\ldots{}it [the word] is really just uncommon for me\ldots{}I just thought about what could commonly be in settings and tried to match the words I see to that. --P6
\end{quote}

 Participants relied on the hovering suggestions (P11), went to the English translation to find the location of an equivalent link (P10, P12), or used their phone's translation feature (P1) to see the interface in English. Our participants also described that they ``would have preferred if it was described as a phrase instead of a single, uncommon word.'' In Amharic Wikipedia, there was one link titled `Minchige', a word that puzzled three of our participants (P2, P3, P10). Only P3, who is a history professor was able to decipher what it means: 

\begin{quote}
    Minchige?\ldots{}ohh maybe it is like `a place for sources'? `-ge', as an extension, is used to say `a place of'\ldots{} some say it is Ge'ez, others say it is Guraginya. For example, Gurage\footnote{name of an ethnic group in Ethiopia} means `a place of the Gura people'. So maybe that is what it is saying.''
\end{quote}
 
The second issue we observed with the Wikipedia interface was how the interface itself is a mix of English and the language of the Wikipedia. Our participants (P2, P3, P6, P7, P10, P11) note this as they interact with the interface: ``it is showing me terms of use in English. What if I cannot read English?'' Mix of languages was not limited to natural language: P3 who attempted to add an image to her article said she is ``not used to this type of language [Wikitext].'' and that the source-based editing feature was ``confusing.''

The third issue we observed was the lack of contained information on how to add new articles, how to add media, or how to recover from errors. Participants would sometimes have to go out of Wikipedia to find information on how to perform certain actions. For instance, after searching for ``How-tos'' on Google and going through some iterations on copyright issues, P3 gave up trying to add the image. Some participants (P3, P6, P10, P12, P13) also had a hard time trying to figure out how to add a new article. Some participants (P3, P10, P13) navigated this by searching for the title they wanted to write about in the Wiki search bar, and clicking the `create new article' link when it returned there were no articles in that title. Two of our participants (P2, P10) were also ``confused'' by the lack of immediate feedback after performing an action; for instance, after pushing publish or adding a footnote. P2 said ``Did I edit it for real or is it an experiment? I am not sure\ldots{}''. Another observation we had about the Wikipedia interface itself is related to how the interface does not allow users to easily recover from errors in low-resourced languages. P10 wrote an article and wanted to connect it to another Wikipedia article. When he searched in Amharic, he could not find the article he wanted to link. He searched in English Wikipedia and found the equivalent, so he decided to create an Amharic article about the same topic. He then created the article in Amharic Wikipedia but made a mistake and used an English title. Once the article was created, he could not figure out how to edit the title. After repeated failed attempts, including trying to use the `Translate this Page' feature, he gave up and just linked the Amharic article with the English title to his original article. Then, ``out of curiosity'', he went to English Wikipedia and tried to create an article with an Amharic title. The interface gave him a notice that he cannot create an article with ``Ethiopic'' script on English Wikipedia: 

\begin{quote}
    By accident if you do something, you cannot change it. Why would it even let me put an English title?\ldots{}You see? In Amharic Wikipedia, I can write in English but in English Wikipedia, I cannot write in Amharic.--P10
\end{quote}

\begin{figure*}[ht!]
    \centering
    \includegraphics[width=\textwidth]{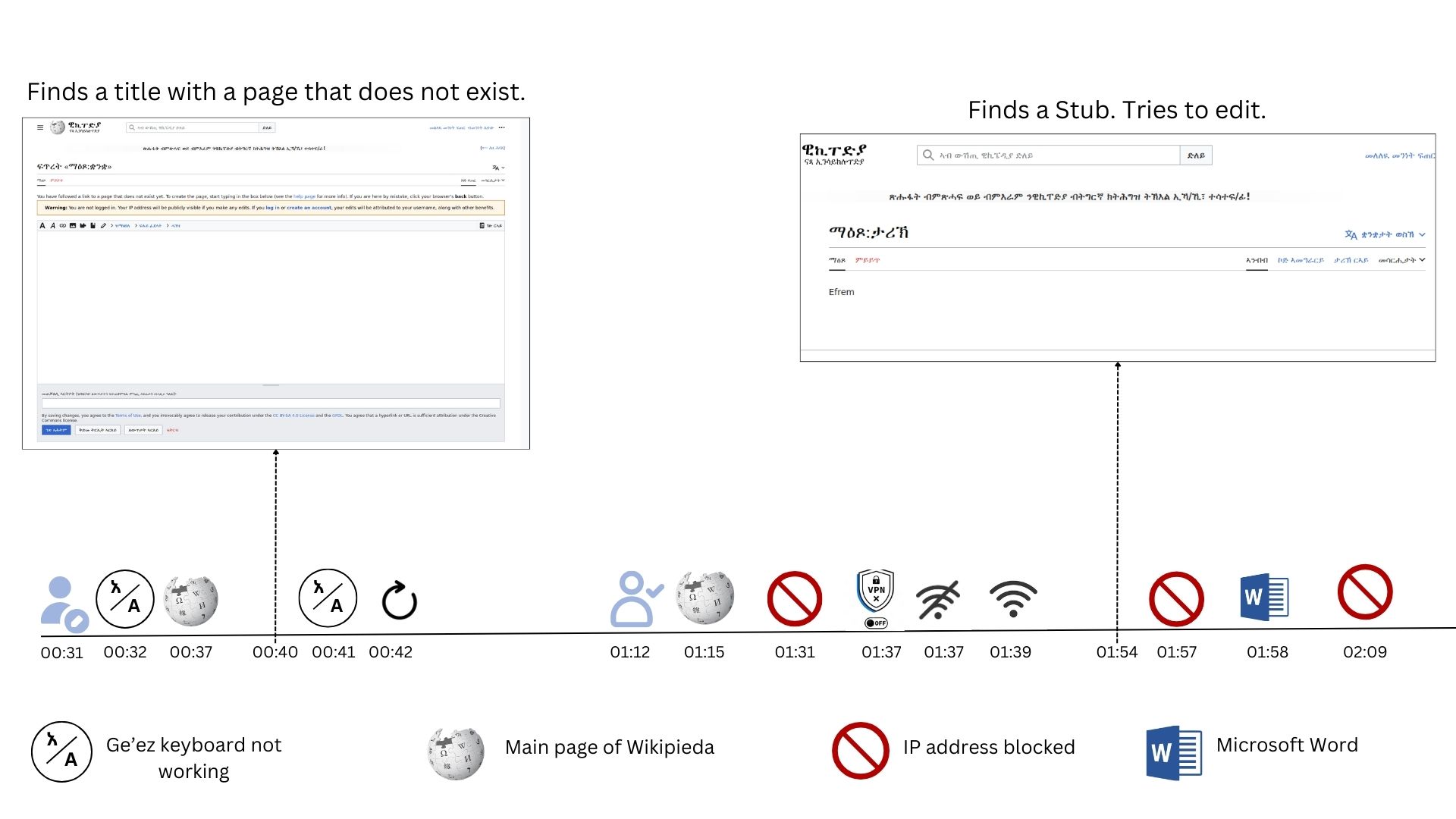}
    \caption{\textbf{P9's timeline showing challenges he faced as he was trying to create an article in Tigrinya.} He first tries to create an account in Tigrinya Wikipedia and notices the keyboard is in Latin. Once he has an account, he follows one of the links on the main page to find an article to edit but finds a ``Page not found'' notice. He tried to create a new page in that category but could not get the Tigrinya keyboard to work on Wikipedia. He then decides to restart his computer, which takes half an hour before he can get back to log in on Wikipedia. Now, when he tries to create a new page for the category, he is told his IP address is blocked. Then, he disconnects his VPN which also ends up disconnecting him from the internet. Once he reconnects, he tries to find a different category to edit and finds a \textit{stub} article. He tries to edit the stub but gets a notice that his IP address is still blocked. At this point, he decided to write the article in Microsoft Word and copy it to Wikipedia. After writing his article, he tries to copy on Wikipedia but is still told his IP is blocked. P9 gives up at this point, saving his article on his local computer and saying he will try again some other time. (see Fig. \ref{fig:p9_screen} in Appendix \ref{screenshots} for larger screenshots)}
    \label{fig:timeline}
\end{figure*}

\subsubsection{Issues with Existing Articles} \label{contextual_quality_issues} We observed that the issues with existing articles (Section \ref{background}) affected the experience of our participants as they attempted to write articles in the respective Wikipedia. We observed that spelling issues affect participants' (P0, P1, P7, P11, P13) writing experience. As P13 explained, he ``goes with his gut'' to correct the words he feels are misspelled. Additionally, participants (P0, P4, P9) note that differences due to dialects result in different spellings of the same word, which as discussed in Section \ref{background}, could trickle down to search experience. Our participants sometimes corrected existing issues, even if that was not their primary goal. For instance, P2 was searching for a city in Ethiopia but could not find it on Wikipedia. After some digging, he found out the Amharic Wikipedia article had an old name of the city and made a note to ``come back and edit this later.'' Similarly, P3 shared how she ``sometimes goes to Wiki[pedia] to find sources and ends up correcting the article.''

 \section{Limitations} \label{limitation} 
Our study focuses on three languages spoken in Ethiopia. While we made a conscious choice to focus on these three languages, we acknowledge that this leads to less focus on any one language.  With a narrow focus on just one, it is possible we could have delved deeper into the challenges in that language. However, our goal was to observe challenges that transcend language boundaries. We note that the languages are not representative of all low-resourced languages, and our study is limited to the Ethiopian context. Future work can explore whether these challenges generalize to other languages and other cultures. Additionally, our second study was with novice contributors, hence, some of the challenges observed could be due to unfamiliarity with the platform. However, in a space where there is already a limited number of contributors, it is crucial to understand the barriers to entry. Additionally, our forum analysis gives insight into the problems faced by contributors who are already active on the platform. 

We focused our study on Wikipedia; hence, the challenges may not generalize to other platforms. We chose Wikipedia as a primary focus because it is considered a mainstream OKR within much of academic computer science, and because it has been the subject of prior works \cite{gallert_indigenous_2016, van_pinxteren_african_2017}. Despite the academic perception that Wikipedia is mainstream and that having large volumes of information available on Wikipedia is positive, these languages may be underrepresented on Wikipedia because communities do not want their data online, or do not want their data specifically on Wikipedia. Our goal with this work is not to argue that communities should or should not put their data on Wikipedia.
Rather, we aimed to investigate the challenges contributors face when they try. As we discuss in Section \ref{intro}, we argue that communities should have that choice---our design shortcomings should not make the choices for them. 


\section{Opportunities for Future Work} \label{lessons}

   Below, we use our findings to ground recommendations for improving future tools, both within Wikipedia and with other language technologies that participants use for contributing to Wikipedia.  

   \paragraph{\textbf{Wikipedia Interface:}} 
   Wikipedia in our target languages includes high rates of dialect-related spelling variations (Section \ref{contextual_quality_issues}) and mixed-language articles (Section \ref{misspellings}). 
   For languages in which different dialects use different spellings, variations in spellings could be handled more explicitly by requiring contributors to specify the target dialect and making the dialect visible during editing interactions. 
   Wikipedia could also introduce tooling to their Suggested Edits App~\cite{ContributorstoWikimediaprojects2023Nov} that automatically flags articles with mixed languages using Language ID \cite{gaim_geezswitch_nodate} technologies. It is important to note that sometimes contributors may want articles to mix languages, for instance when describing cognates in different languages (e.g, \cite{ContributorstoWikimediaprojects2023Nov}). To avoid removing such articles, we can leave the decision to the editors instead of automatically removing the article. 
   Interventions like adding features to show what percentage is in what language and highlighting the different languages with different colors could also support editors in their decision-making.
   We also observed a contributor supported the Amharic Wikipedia by suggesting articles from English Wikipedia that discussed Ethiopian contexts (Section \ref{forum_findings}). This kind of activity has the potential to aid progress on the lack of relevance trap discussed in Section \ref{related_work} and could be supported via an automatic tool.

     
  \paragraph{\textbf{Information Retrieval:}}  We observed that contributors suffer from the poor performance of Information Retrieval systems both within (Section \ref{socio_findings} and \ref{misspellings}) and outside (Section \ref{socio_findings}) of the Wikipedia platform. We observed that contributors faced challenges arising from (1) different spellings of the same word (Section \ref{background} )and (2) failure to retrieve content they knew existed.  
    Previous work in NLP has used phonetic features to improve search in other contexts (e.g., E-commerce sites dealing with misspellings~\cite{Yang2022May}). Platforms could use similar approaches for improving Information Retrieval performance in languages with multiple spellings for words.
    
   \paragraph{\textbf{Machine Translation:}} We observed that participants used Machine Translation to create new articles and to find translations or spellings of isolated words (see Section \ref{lang_support}). In our forum analysis, we observed that forum posters suggested using translation systems to relieve the challenges of typing in non-Latin scripts (Section \ref{forum_translation}). We found that Machine Translation tools (1) fail to adopt the correct calendars in translation, (2) (still) exhibit gender bias and output unrelated and violent content, and (3) are too `literal' and lack cultural context (Section \ref{lang_support}). HCI researchers could build tools for coupling Machine Translation systems with tools to handle structured elements of the transformation process---such as transforming to different calendar systems. Researchers could adopt methods from, for instance, previous work~\cite{Kim2016May} in contextualizing spatial measurements to users. NLP researchers could also proactively mitigate biases in Machine Translation systems by, for instance, using strategies from previous work~\cite{Stafanovics2020Nov} and explicitly specifying how socially and culturally mediated features like gender and religion are described in different low-resourced languages. An alternative could also be to use back-translation~\cite{Shigenobu2007} to show the author potential deviations from their input text. 

\paragraph{\textbf{Input Modalities:}} 
Although there are multiple Ge'ez-specialized keyboards~\cite{fyngeez_2023, keyman_2023, Metaferia2023Dec}, our findings indicate there are situations where their design choices lead to negative experiences (Section \ref{lang_support}). 
We propose a potential intervention: designers can offer interaction paradigms built around speech technologies in place of typing. 
Granted, speech technologies are not yet performing well enough for low-resourced languages~\cite{Robinson2022Jul, Ogayo2022Jul}. 
However, small-scale, specialized applications of speech technologies have the potential to aid low-resourced language speakers. 
Previous work~\cite{Reitmaier2022Apr} improved speech recognition systems for WhatsApp voice messaging usage in two low-resourced languages using participatory design approaches. 
HCI and NLP researchers could adopt similar participatory methods that center the needs of OKR contributors.
On a related note, our participants indicated how the purely text-based interaction paradigm results in the loss of culture and stories, especially as it relates to community members who are marginalized (Section \ref{lang_support}). 
Interface designers could incorporate multi-modal interaction paradigms (e.g. \cite{patel_ao_2010}) to increase inclusivity instead of requiring communities to adopt Western defaults.  

\section{Discussion} \label{disscussion}


In Section \ref{lessons}, we provided concrete suggestions for improving tools and interfaces based on our findings. We base our recommendations on the observation that challenges faced by low-resourced language contributors lie at the intersection of social, technical, and cultural dimensions. Hence, solutions to address the challenges (1) require deeper reflections on how to embed cultural contexts into technological interactions at different stages in the writing process; (2) involve multiple, interacting technologies; and (3) need to ensure community members retain their agency. We hope our recommendations will guide platforms, designers, and researchers in building tools that reflect community practices. 

 As we showed in Section \ref{socio_findings} and Section \ref{forum_requests}, low-resourced language contributors face challenges related to low-resourcedness, which do not always stem from technology design decisions, although they may be heightened by those decisions. In line with previous work~\cite{nabyonga-orem_article_2020}, we observed that financial barriers prevent low-resourced language contributors from creating and accessing scholarly work. We also observed that the lack of reference materials in local languages and within local contexts results in one-sided publications about these communities being the easiest to find. Previous work \cite{khatri_social_2022}, found that offering monetary resources had led to more territorial and strict policy enforcement by Wiki admins and caution for a more thorough reflection of existing power structures in the communities before introducing such interventions. Our findings indicate that financial barriers are not limited to direct interactions with Wikipedia. Financial barriers are also preventing access to scholarly resources. In response, we echo decolonial scholars' calls to support platforms like AfricArXiv \cite{noauthor_africarxiv_nodate} where African researchers can publish and access research in local languages. Further, we showed in Section \ref{socio_findings} how standard cyber-security efforts, like IP blocking, hamper contributors' efforts to circumvent challenges caused by political and societal problems. Platforms, such as OKRs, should have these challenges in mind when they design and implement mechanisms to protect their users.


Designing for communities that deviate from the `default' is not always straightforward. Previous works have shown that designing for low-resourced language-speaking groups presents barriers due to difficulty accessing experts \cite{yimam_exploring_2020} and difficulty building trust \cite{abebe_narratives_2021, donavyn_maori_nodate}. Language barriers and lack of access to technology might also impact the ability of researchers to engage with such communities. Additionally, as we discuss in Section \ref{related_work_harms}, technological solutions may also expose communities to exploitation and harm. Previous works~\cite{cajamarca_co-design_2022, chowdhury_co-designing_2023, le_dantec_strangers_2015, harrington_engaging_2019} show examples of how using participatory design~\cite{muller_participatory_1993} and action research~\cite{lewin_action_1946} can center community values and emphasize the decision making of community members. Further, we want to emphasize that research does not stop at publication; we have seen in our work how communities find it difficult to find publications in their own languages. We plan to take our own advice and translate our work into the three languages of our study and make it available to community members via blog posts and social media posts. We hope future research with low-resourced language communities follows the same approach.

\section{Conclusion} \label{conclusion}
We investigated the challenges low-resourced language contributors face when interacting with online knowledge repositories. We analyzed forum discussions from Talk Pages in Wikipedia in three languages: Amharic, Afan Oromo, and Tigrinya. We also conducted a contextual inquiry study with 14 novice low-resourced language contributors. We then discussed the implications of our work. By centering cultural aspects in the design and development of technologies that serve global, diverse communities, we can build inclusive platforms. We believe this work provides insights to researchers, designers, and platform managers in building inclusive technologies, especially as it relates to low-resourced language speakers. 
\begin{acks}
This work is supported in part by NSF grants FW-HTF 2129008, CA-HDR 1936731, and CA-HDR 2033558, as well as by gifts from Google, Microsoft and, and Sigma Computing. Sarah E. Chasins is a Chan Zuckerberg Biohub Investigator. Hellina Hailu Nigatu is a SIGHPC Computational and Data Science Fellow. We thank our colleagues from WhoseKnowledge?, Lesan AI, and Distributed AI Research Institute (DAIR) for early feedback on this work. We thank Chris Emezue from Masakhane for his early feedback and conversations. We would also like to thank members and friends of PLAIT lab for their feedback on this work. Additionally, we thank reviewers from CHI'2024 and from AfriCHI'2023. We thank Bontu Fufa Balcha who helped us translate some of the Afan Oromo text. Finally, to our participants, we say: Enamesegnalen. Yeqeniyelna. Galatoomaa. 
\end{acks}

\bibliographystyle{ACM-Reference-Format}
\bibliography{references, custom} 

\appendix
\section{Screenshots from Contextual Inquiry Sessions} \label{screenshots}
\begin{figure*}[ht!]
    \centering

    \subfloat[Page does not exist notice when participants follow the link for `Language'.]{\includegraphics[width=0.8\textwidth]{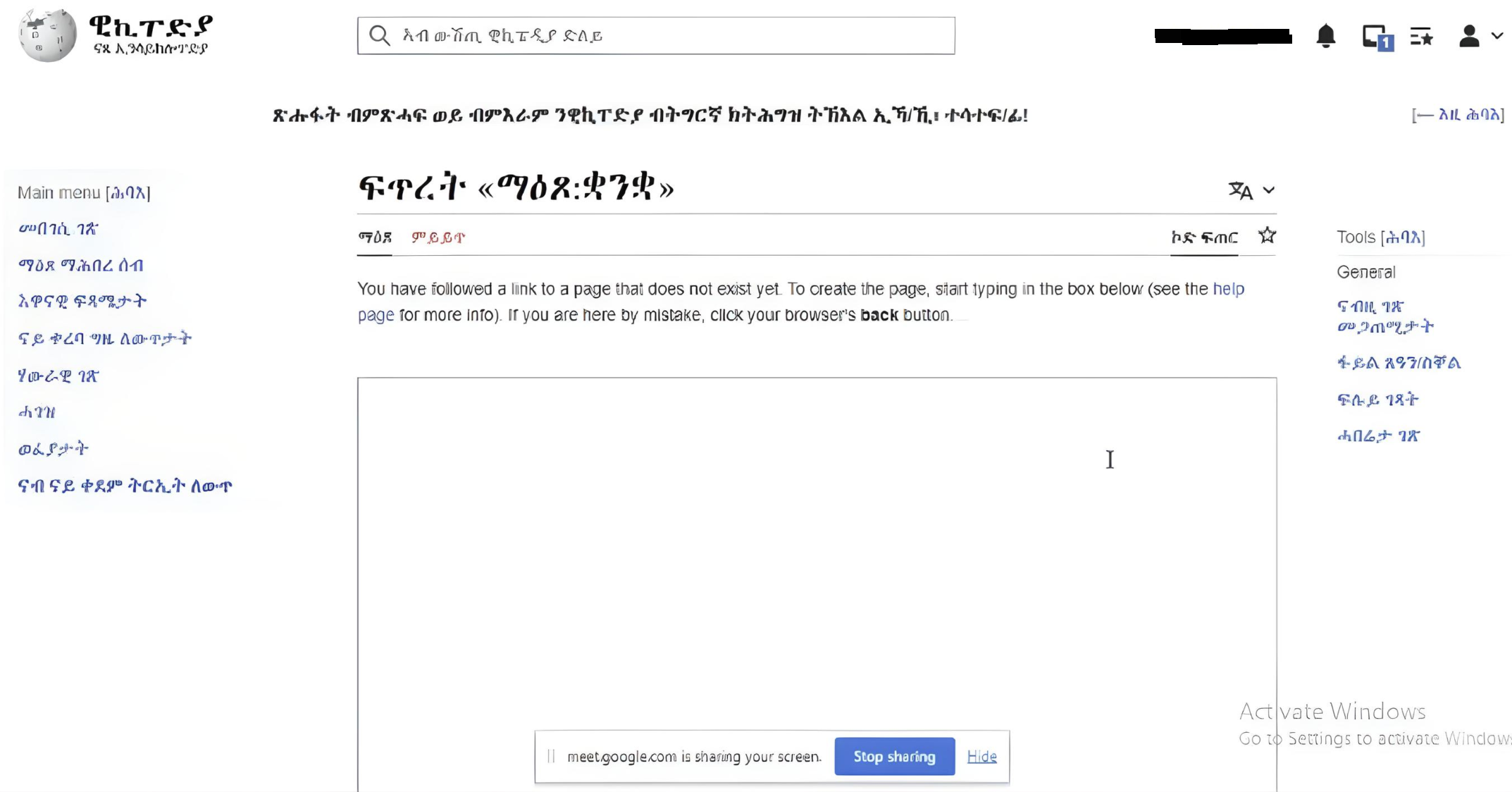}\label{fig:page_does_not_exist}} 

    \subfloat[IP is blocked notice.]{ \includegraphics[width=0.8\textwidth]{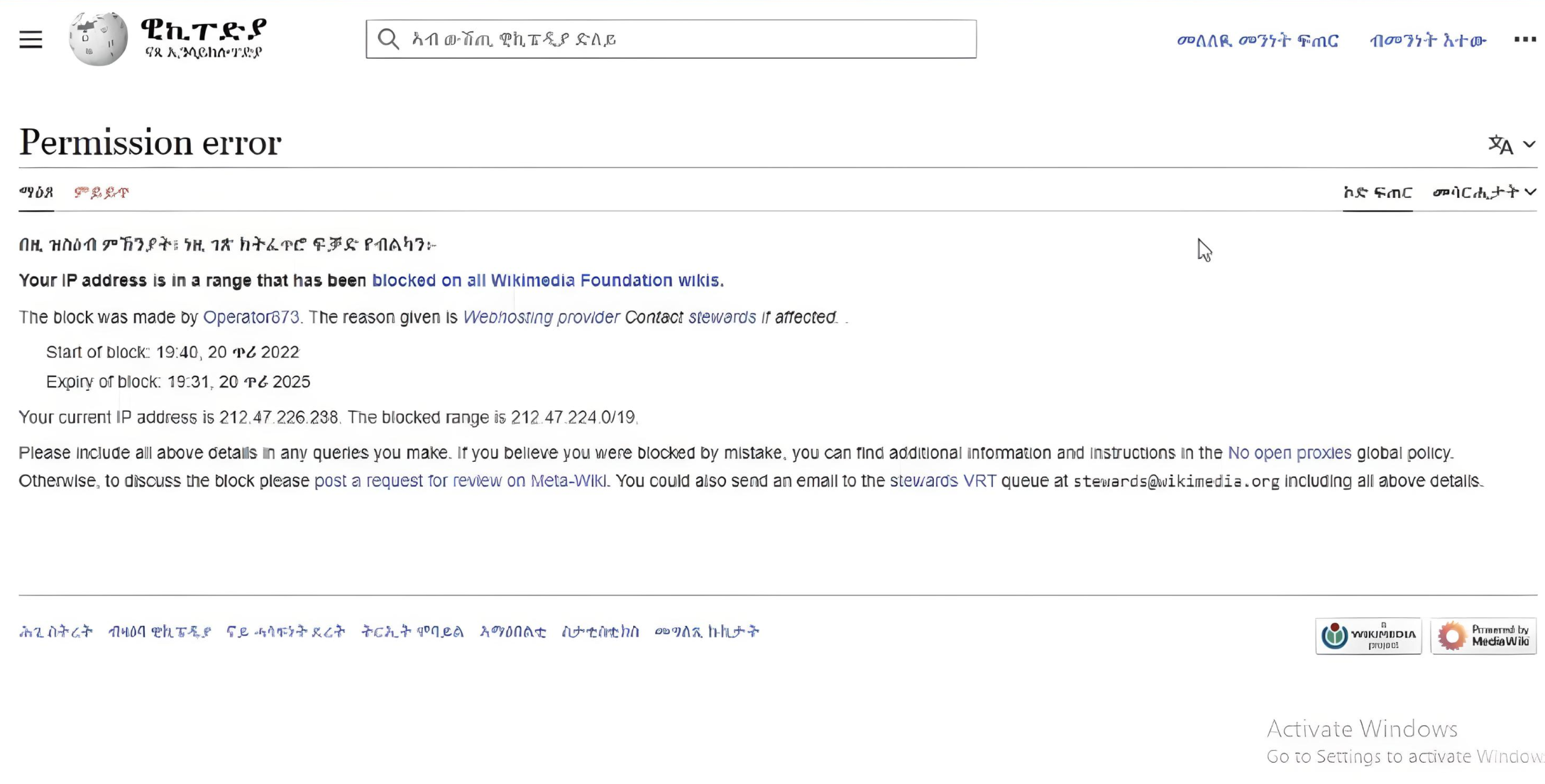}\label{fig:IP_blocked}
          }
    \caption{\textbf{Screenshots from the session with P9 and P4.} In Fig. \ref{fig:page_does_not_exist}, P9 and P4 followed the link to ``qwanqwa'' which means `language' from the main page and ended up with the interface above, telling them the page does not exist and allowing them to edit it. We have redacted the username of our participant in the top right corner. Note the interface contains a mix of English and Tigrinya. In Fig. \ref{fig:IP_blocked} P9 received a notice showing that their IP is blocked. In this case, the error message in Tigrinya says ``You don't have permission to create this page due to the following reasons:'' and continues to state the error in English. Wikipedia policy \cite{noauthor_helpi_2023} states that reasons for being blocked could be ``Using a VPN or other anonymizing proxy service''. The policy further states that one can submit an appeal or request IP block exemption.}
    \label{fig:p9_screen}
\end{figure*}

In this section, we provide screenshots from our contextual inquiry sessions with participants in Study 2. 
 Fig. \ref{fig:p9_screen} shows screenshots from P9's session where they got a `page does not exist' notice following a link from the main page and a Permission error when using VPN. 
    

\section{Coding Scheme and Generated Codes} \label{apd:Coding}
In this section, we provide the themes from our two studies and give examples of our open codes. In Section \ref{study_1_codes}, we give the themes for Study 1 and in Section \ref{hierarchy_of_themes}, we give the themes and codes from our analysis for Study 2. 
\subsection{Open Codes and Themes from Analysis of Study 1 (Wikipedia Forum Data)} \label{study_1_codes}
In this section, we provide Table \ref{tab:codes} with the second- and first-level themes as well as some example open codes from our analysis of forum data from the three Wikipedias. In total, we had 6 Themes, 24 sub-themes, and 265 open codes.

\begin{table*}[h!]
\small
\centering
\begin{tabular}{p{3cm}|p{4cm}|p{8cm}}
\toprule
\textbf{Second-Level Theme}& \textbf{First-Level Themes} & \textbf{Example open-codes}\\
\midrule
\multirow{7}{3cm}{Engagement among Contributors (96)}& Advertising (14) & asking carrier groups and Addis Ababa University library people; proposing for expansion using external tools like Paltalk\\ \cline{2-3}
 & Division of Roles (19) & proposing to divide work based on the people's professions; against division since the wiki is too small\\ \cline{2-3}
  & Words of Encouragement (16) & encouraging growth; appreciation of contributors\\ \cline{2-3}
   & Security and Vandalism (9) & coordinating on opening the locked page; there is no controller/admin \\ \cline{2-3}
    & Statistical Goal Setting (13) & predicting future article count based on growth rate; comparing to article numbers in other languages\\ \cline{2-3}
     & Seeking Guidance (15) & asking information on how to set font; seeking support for edits as a non-native speaker\\ \cline{2-3}
      & Wikipedia Bureaucratic Content (10)  & request to review code of conduct; explain election process\\ \cline{1-3}
\multirow{4}{3cm}{Issues with (Wikipedia) Design (54)}& Font (25) & expressing font display issues; discussion by someone who cannot read the language; asking for help to write in their own script in another language\\ \cline{2-3}
 & Input Mechanism (11) & expressing the difficulty of using Latin keyboard; noting comments of time-consuming issue of typing in Amharic.\\ \cline{2-3}
   & Visual changes (12) & suggesting using other languages' layout as a template; calling for more structure in main page; suggesting color alternatives for `bright' colors\\ \cline{2-3}
    & Broken Links (6) & indicating an article is not connected to its English page; asking why the article they previously read is no longer there\\ \cline{1-3}
 \multirow{2}{3cm}{Common Requests (42) }& New Articles (21) & article request for school in DC; request to start new article from an English wiki \\ \cline{2-3}
 & Entity Extraction (8) & request to extract the name from a file in Amharic; request to search for the names in Tigrinya and Amharic in another file\\ \cline{2-3}
  & Funding (6) & request support for mobile airtime top-up; questioning how much financial support their is for Amharic Wiki\\ \cline{2-3}
  & Media Requests (7) & asking for a photo of the airport; request from diaspora to locals to take pictures\\ \cline{1-3}
   
 \multirow{3}{3cm}{Relations across Languages (39) }& non-Local Translations (10)& asking to add an article about Polish Prime Minister in Afan Oromo; requesting translation of Amharic letter for English Wikipedia\\ \cline{2-3}
 & Linking and Edits (15) & asking to add a link to another language Wiki to the main page; asking to correct a mistake about their language\\ \cline{2-3}
  & Starting New Wikipedia (14) & suggest Ge'ez should be independent; practices of other Wiki of ancient and modern languages\\ \cline{1-3}

\multirow{3}{3cm}{(Use of) Translation (23)}& Translation as alternative  (7) & asking for translation during spare time to increase content; suggesting using translation to overcome typing challenges\\ \cline{2-3}
 & Translating Interface Words (11) & asking for help to translate interface words; asking for translations to be simple for interface words\\ \cline{2-3}
  & Translating Scientific Terms (5) & especially concerned with terms related to science; propose using the translatewiki platform for the collaborative translation\\ \cline{1-3}

\multirow{2}{3cm}{Quality Issues with Existing Articles (11)}& Spelling (5) & created duplicate topics and found out it exists in different spelling; fear misspelling will hinder the wiki unusable\\ \cline{2-3}
 & Short articles/Stubs (3) & indicating new contributors create new articles with no body; complaining about stubs\\ \cline{2-3}
 
 & Language Mixing (3) & complaining wiki is unusable; indicating most content is in English\\ \cline{1-3}

 \hline
\end{tabular}
\caption{\textbf{Second- and first-level themes from our forum analysis.} We provide the number of unique open codes that support each second- and first-level theme in brackets, ordering them from the highest to lowest number of unique codes per the second-level theme. `Engagement among Contributors' was the highest supported theme followed by 'Issues with (Wikipieda) Design'.}
\label{tab:codes}
\end{table*}

\subsection{Open Codes and Themes for Study 2 (Contextual Inquiry with Novice Contributors)} \label{hierarchy_of_themes}
In Table \ref{tab:study_2_codes}, we provide the hierarchical themes from our contextual inquiry study and provide example open codes. For this study, we had three levels of themes and 443 unique open codes. 

\begin{table*}[h!]
\small
\centering
\begin{tabular}{p{1.5cm}|p{2cm}|p{5cm}|p{6cm}}
\toprule
\textbf{Third-Level Theme} & \textbf{Second-Level Theme} & \textbf{First-Level Theme} & \textbf{Example Open Codes}\\
\midrule
\multirow{2}{1.5cm}{Issues with Language Support Technology (164)} & \multirow{4}{2cm}{Input Mechanisms (62)} & Oral traditions and lack of voice interface support (9) & talking about to communities that predominantly use oral traditions
\\ \cline{3-4}
& & Using external Software (12) & writes in Microsoft Word
\\ \cline{3-4}
& & Variation in the device used for writing (10) & indicating preference for writing on their phone 
\\ \cline{3-4}
& & Variation and challenges related to keyboards (31) & multiple keystrokes for one letter
\\ 
\cline{2-4}

& \multirow{5}{2cm}{Translation Issues (102) } & Translation of scientific terms (11) & 
transliterates scientific terms \\ \cline{3-4}
& & Syntax issues and direct translations (7) & translation follows syntax of target language; 
\\ \cline{3-4}
& & Bias and the role of cultural context (42) & output says `father' for a women organization that have `mother'\\ \cline{3-4}
& & Post and Pre-editing (13) &
removes mixed English content in output  \\ \cline{3-4}
& & Contexts of Use of Human and Machine Translation Systems (29) & 
translates a full paragraph with Google Translate \\ 
\cline{1-4}

\multirow{2}{1.5cm}{Socio-Political issues (114)} & \multirow{7}{2cm}{(Lack of) Availability of Scholarly Resources (103)} & Financial barriers to scholarly work and publication (7) & cannot access pay-walled books and publications
\\ \cline{3-4}
& & Conducting research in local languages (11) & explaining how data collection for research is done in local languages; 
\\ \cline{3-4}
& & Available resources with Western gaze (14) & 
there is a (semantic) mistake in the European source being used \\ \cline{3-4}
& & Social Media influence of political content in the search experience (28) & 
gets political posts for searching [ethnic group] culture \\ \cline{3-4}
& & Local works being locked in hard-copy materials (9) & format for adding PDF citation requires extra steps;
\\ \cline{3-4}
& & How users deal with lack of resources (26) & give up the topic after failed attempts to find a reference; 
\\ \cline{3-4}
& & Biased, one-sided views even in local materials (8) & 
cannot find any positive resources about a Muslim leader \\ 
\cline{2-4}

&  IP Blockage (11) & VPN service leading to IP Blockage (11) & 
gets disconnected from internet\\ \cline{1-4}

\multirow{2}{1.5cm}{Issues with Wikipedia Interface (95)} & \multirow{5}{2cm}{Understanding Interface Words (48)} & Translation, transliteration, and English alternatives (13)  & used the phones translate option to English 
\\ \cline{3-4}
 & & Not ‘Everyday use’ words (12) & comments it would make more sense as a sentence; cannot figure out which button is for publishing 
 \\ \cline{3-4}
 & &  Differences in  dialects (5) & explaining in some dialects it makes sense but in others not 
 \\ \cline{3-4}
 & &  Grammar, spelling, syntax issues (7)& long vs short vowels that change meaning due to spelling differences 
 \\ \cline{3-4}
 & &  Interface in a mix of English and local language (11) & interface to add citation is a mix of English and Amharic
 \\ \cline{2-4}
 
 & \multirow{6}{2cm}{Lack of contained information and feedback (47) } & Lack of information on adding media (4) & unsure what file format (JPEG, PNG, JPG) made the task successful
 \\ \cline{3-4}
 & & Lack of information on proper attribution and citation (10) & fearful of republishing someone else's work as their own
 \\ \cline{3-4}
   & & Lack of information on creating new articles (14) & 
   googles how to add a new article on Wikipedia\\ \cline{3-4}
  & & No immediate feedback on the effect of change (4) & unsure if the article was published or is only reflecting on their end; cannot see the citation they added\\ \cline{3-4}
   & & Unable to reverse the committed change (5) & accidentally created article with English title; 
   \\ \cline{3-4}
  & & Disparate treatment in English vs Local Wikipedia (7) & 
  gets notice Ethiopic script is not allowed
  \\ \cline{3-4}
  
    & & No of summary of article(s) (3) & 
    fears edit might duplicate information in the article\\ \cline{1-4}
    
\multirow{2}{1.5cm}{Problems with Existing Articles (70)} & 

\multirow{3}{2cm}{Lack of cultural relevance (20) } & Issues with main page article (3) & 
asking why Big Mac is the first article they see \\ \cline{3-4}

& & Content unrelated to the title (3) & indicates the content is a political post \\ \cline{3-4}
& & Searching for African content (14) & indicates frustration with lack of African content
\\ \cline {2-4}

& Problems with article length (9) & Stubs and one sentence articles (9) & 
article stops in the middle of a sentence\\ \cline{2-4}

& \multirow{2}{3cm}{Quality issues (41)} & Articles with outdated information (11) & old data about population;
\\ \cline{3-4}
& & Articles with wrong links (14)& link is to a different historical figure who has the same name;
\\ \cline{3-4}
& & Articles with spelling issues (16) & 
edits spelling error on an existing article  \\ \cline{1-4}

\end{tabular}
\caption{\textbf{Hierarchical themes from our contextual inquiry study.} For hierarchical themes, we provide the number of unique open codes that support the theme in brackets, ordering them from the highest to lowest number of unique codes per third-level theme. We find that `Issues with Language Support Technology' is the most supported theme. }
\label{tab:study_2_codes}
\end{table*}

\end{document}